\documentclass[sigconf]{acmart}

\usepackage{amssymb}

\usepackage[english]{babel}
\usepackage{blindtext}

\renewcommand\footnotetextcopyrightpermission[1]{} %
\setcopyright{none}

\acmDOI{}

\acmISBN{}

\acmConference[]%
\acmYear{}%

\acmPrice{}

\usepackage{tikz}
\usepackage{amsmath, nccmath}
\usepackage[english]{babel}
\usepackage{blindtext}
\usepackage{outlines}
\usepackage{graphicx}
\usepackage{times}
\usepackage{hyperref}
\usepackage{ulem}
\usepackage{xspace}
\usepackage[dvipsnames]{xcolor}
\usepackage{balance}
\usepackage{outlines}
\usepackage{tikz}
\usepackage{caption}
\usepackage{multirow}
\usepackage[]{url}
\usepackage{todonotes}
\usepackage[compact,small]{titlesec}
\usepackage{algorithm}
\usepackage{algpseudocode}
\usepackage{tabularx}
\usepackage{array}
\usepackage{listings}
\usepackage[font=small,labelfont=bf]{caption}
\usepackage{xurl}
\usepackage[capitalise]{cleveref}
\usepackage[inline,shortlabels]{enumitem}
\setitemize{noitemsep,topsep=0pt,parsep=0pt,partopsep=0pt,leftmargin=1em}
\usepackage{makecell}
\usepackage{multicol}
\usepackage{subcaption}

\usepackage[table]{xcolor} %

\usepackage[font={small,bf},labelfont={small,bf},textfont={small},belowskip=-5pt]{caption}
\usepackage[font={small,bf},labelfont={small,bf},textfont={small},belowskip=-6pt,aboveskip=1pt, labelformat=simple]{subcaption}

\usepackage[subtle]{savetrees}

\newcommand{\ie}{\emph{i.e.,}\xspace}
\newcommand{\eg}{\emph{e.g.,}\xspace}
\newcommand{\etc}{\emph{etc.}\xspace}
\newcommand{\vs}{\emph{vs.}\xspace}

\usepackage[english]{babel}
\usepackage{blindtext}

\usepackage{amsmath}
\usepackage{caption}
\usepackage{graphicx}
\usepackage{xurl}
	\newcolumntype{C}[1]{>{\centering\let\newline\\\arraybackslash\hspace{0pt}}m{#1}}

\newcommand{\bl}{\begin{enumerate*}[(1)]}
\newcommand{\el}{\end{enumerate*}}

\usepackage{pifont} %

\crefname{equation}{Eq.}{Eqs.}
\Crefname{equation}{Eq.}{Eqs.}
\crefname{figure}{Fig.}{Figs.}
\Crefname{figure}{Figure}{Figures}
\crefname{table}{Table}{Tables}
\Crefname{table}{Table}{Tables}
\crefname{property}{Property}{Properties}
\Crefname{property}{Property}{Properties}
\let\oldappendix\appendix
\renewcommand{\appendix}{%
  \oldappendix%
  \crefalias{section}{appendix}%
}

\AtBeginDocument{%
    \crefname{appendix}{}{}%
    \Crefname{appendix}{}{}%
    \crefname{section}{}{}%
    \Crefname{section}{}{}%
    
    \crefformat{appendix}{Appendix~#2#1#3}
    \Crefformat{appendix}{Appendix~#2#1#3}
    \crefrangeformat{appendix}{Appendices~#3#1#4--#5#2#6}
    \Crefrangeformat{appendix}{Appendices~#3#1#4--#5#2#6}
    \crefmultiformat{appendix}{Appendices~#2#1#3}{ and~#2#1#3}{, #2#1#3}{ and~#2#1#3}
    \Crefmultiformat{appendix}{Appendices~#2#1#3}{ and~#2#1#3}{, #2#1#3}{ and~#2#1#3}

    \crefformat{section}{\S#2#1#3}
    \Crefformat{section}{\S#2#1#3}
    \crefrangeformat{section}{\S\S#3#1#4--#5#2#6}
    \Crefrangeformat{section}{\S\S#3#1#4--#5#2#6}
    \crefmultiformat{section}{\S\S#2#1#3}{ and~#2#1#3}{, #2#1#3}{ and~#2#1#3}
    \Crefmultiformat{section}{\S\S#2#1#3}{ and~#2#1#3}{, #2#1#3}{ and~#2#1#3}
}

\newcommand{\thm}[1]{Thm.~\ref{#1}} %
\newcommand{\inv}[1]{Inv.~\ref{#1}} %

\newcommand{\be}{\begin{equation}}
\newcommand{\ee}{\end{equation}}

\providecommand{\vs}{{vs.}\xspace}
\providecommand{\ie}{{i.e.,}\xspace}
\providecommand{\eg}{{e.g.,}\xspace}

\providecommand{\etc}{{etc.}\xspace}

\providecommand{\sysname}{{Ofan}\xspace}

\newcommand{\T}[1]{\par\noindent\textbf{#1}} %
\newcommand{\mypar}[1]{\T{#1.\xspace}}

\newtheoremstyle{boldthm}{}{}{\itshape}{}{\bfseries}{.}{ }{\thmname{#1}\thmnumber{ #2}\thmnote{ (#3)}} %
\theoremstyle{boldthm}
\newtheorem{invariant}{Invariant}
\newtheorem{theorem}{Theorem}%
\newtheorem{property}{Property} %

\newcommand{\bp}{\begin{proof}}
\newcommand{\bpo}{ \begin{proof}[Proof Outline] }
\newcommand{\ep}{\end{proof}}       %
\newcommand{\proofof}[1]{\begin{proof}[Proof of #1]} %

\newcommand{\para}[1]{\left( #1 \right)}        %
\newcommand{\brac}[1]{\left\{ #1 \right\}}
\newcommand{\sbrac}[1]{\left[ #1 \right]}
 
\newcommand{\ceil}[1]{\left\lceil #1 \right\rceil} %

\newcommand{\unit}[1]{\;\mathrm{#1}} %
\usepackage{upgreek} %
\newcommand{\us}{\unit{\mu s}}
\newcommand{\p}[1]{\Pr \para{#1}}  %

\renewcommand{\th}{\ensuremath{^\text{th}}}

\newcommand{\vx}{\checkmark\kern-1.1ex\raisebox{.7ex}{\rotatebox[origin=c]{125}{--}}} %

\newcommand{\dr}{DR\xspace} 
\newcommand{\DR}{\dr} 

\newcommand{\htsim}{\textit{htsim}\xspace} 
\newcommand{\lblabel}[1]{\textsc{\small{#1}}\xspace}
\newcommand{\flow}{\lblabel{Flow}} 
\newcommand{\subflow}{\lblabel{Subflow}} 
\newcommand{\flowlet}{\lblabel{Host Flowlet AR}}
\newcommand{\hpkt}{\lblabel{Host Pkt}}
\newcommand{\spkt}{\lblabel{Switch Pkt}}
\newcommand{\hpktar}{\lblabel{Host Pkt AR}}
\newcommand{\spktar}{\lblabel{Switch Pkt AR}}
\newcommand{\jsq}{\lblabel{JSQ}}
\newcommand{\rsq}{\lblabel{RSQ}}
\newcommand{\simplerr}{\lblabel{Simple RR}}
\newcommand{\hdr}{\lblabel{Host DR}}
\newcommand{\sdr}{\lblabel{Switch DR}}
\newcommand{\name}{\lblabel{Ofan}}

\newcommand{\m}{m}
\newcommand{\n}{n}
\newcommand{\C}{C}
\newcommand{\q}{q(\m)}
\renewcommand{\k}{k}
\renewcommand{\a}{a}
\newcommand{\pkt}{L_{\text{PKT}}}

\newcommand{\rhomax}{\rho_{\text{max}}}
\renewcommand{\H}{H}
\newcommand{\Td}{T_d}
\newcommand{\Ta}{T_a}
\newcommand{\Tg}{T_g}
\renewcommand{\d}{T'_d}
\renewcommand{\a}{T'_a}
\renewcommand{\p}{T_p}

\begin{document}

\date{}

\title{\ Load Balancing for AI Training Workloads}

\newcommand{\aut}[2]{#1\texorpdfstring{$^{#2}$}{(#2)}}  %
\author{
  \aut{Sarah McClure}{1} \quad 
  \aut{Evyatar Cohen}{2} \quad 
  \aut{Alexander Shpiner}{3} \quad 
  \aut{Mark Silberstein}{2,3}\\ 
  \aut{Scott Shenker}{1,4} \quad
  \aut{Sylvia Ratnasamy}{1} \quad
  \aut{Isaac Keslassy}{1,2}
}%
\affiliation{
$^1$ \textit{UC Berkeley}\quad 
$^2$ \textit{Technion} \quad 
$^3$ \textit{NVIDIA} \quad 
$^4$ \textit{ICSI} 
}
\renewcommand{\shortauthors}{McClure et al.}

\sloppypar
\begin{abstract}
The extreme bandwidth demands of AI training has made load-balancing a critical component in AI fabrics, and a variety of load-balancing designs have emerged in recent work from both industry and research. However, there is currently little consensus on which design approach dominates or the conditions under which an approach dominates. We also lack an understanding of how far these approaches are from optimal. 

We provide a technical foundation for answering these questions by systematically evaluating leading load-balancing designs, while decoupling them from specific congestion control and loss recovery stacks. 
We find that load-balancing based on packet spraying dominates traditional approaches that load balance traffic at flow, flowlet, or subflow granularities. When comparing host- vs switch-based approaches to packet spraying, we find that they perform similarly in failure-free scenarios but that a host-based approach dominates under link failure because of its rapid visibility into end-to-end path conditions. We also identify that no leading approach achieves optimal $O(1)$ queue scaling at maximum utilization. We demonstrate why a destination-based rotation (DR) discipline can reach this optimum and introduce \name, a switch-based implementation of DR that we show offers valuable performance gains over other packet spraying approaches. 
\end{abstract}

\maketitle

\section{Introduction}
\label{sec:intro}

Historically, TCP's intolerance for packet reordering has necessitated a rigid single-path-per-flow paradigm. While the downsides of this constraint have long been recognized and various alternatives proposed, network solutions have continued to respect this limitation, ranging from standard ECMP to more recent proposals like Meta's enhanced ECMP~\cite{meta} and Google's PLB~\cite{plb}. These approaches improve performance by shifting path assignments, but still pin a flow to a single path at any given time. However, the extreme bandwidth demands of AI workloads are forcing operators to abandon this paradigm in favor of aggressive packet-level spraying.

AI workloads pose extreme requirements on the network to limit the worst-case completion time among flows.
The needs of AI workloads (which we discuss more thoroughly in \S\ref{sec:background}) 
have led to active exploration, mostly in industry, of designs that go beyond the single-path paradigm. 
Here we mention two efforts in particular. The Ultra Ethernet Consortium (UEC) \cite{uec} is considering new transport designs that include packet spraying and packet trimming \cite{ethereal, uec-spec}. Nvidia has developed Spectrum-X which uses a switch-based packet-spraying approach and RDMA NICs that accommodate out-of-order delivery \cite{spectrumx, nvidia-ooo}. These efforts suggest that we are finally seeing a broad appetite in industry to consider load balancing techniques that no longer restrict packet flows to a single path.

Still, we were surprised to find that many basic design questions remain open.
For instance, while it seems clear that packet spraying is a robust technique for spreading load along all suitable paths, we were not able, in our literature search, to find answers to such key questions as: 
(i)~How do state-of-the-art load-balancing approaches compare when not coupled with a particular network stack?
(ii)~Is it better to implement packet spraying in switches (as in Spectrum-X and pFabric~\cite{pfabric}) or in hosts (as in NDP~\cite{ndp} and REPS~\cite{bonatoreps})? 
(iii)~Do the benefits of (various forms of) packet spraying remain when we introduce failures or the workload creates incast?
(iv) Is failure (and other asymmetry) best handled in the network on in the host?
(v) And how sensitive are performance differences to the nature of the workload and network parameters such as packet size and buffer size?

The ongoing industry efforts suggest there is a lack of consensus on these questions. 
For instance, Nvidia supports switch-based packet spraying; the UEC suggests host-based spraying; Google's Falcon design \cite{falcon} uses host-implemented path adaptation as in PLB~\cite{plb} but does not use packet spraying; 
Meta uses an enhanced ECMP and traffic engineering (TE) \cite{meta}; Alibaba uses oblivious packet spraying \cite{alibaba-stellar} and a host-managed and topology-aware subflow approach \cite{alibaba}; and ByteDance uses  switch-based, congestion-aware global load balancing \cite{bytedance-glb}. 

Beyond the lack of industry consensus on specific implementations, there is currently no rigorous understanding of whether existing load-balancing contenders represent the theoretical limit of performance for AI training. Specifically, it remains unclear how close these current solutions are to an optimal bound. Quantifying this gap is essential; it provides the necessary roadmap to determine whether the community should continue designing novel load-balancing schemes, focus on the incremental optimization of existing ones, or pivot toward entirely new architectural directions.

Answering such questions is challenging because it requires decoupling load balancing from the other aspects of a transport protocol (such as congestion control and loss recovery). Given the competitive and evolving nature of the AI market, we cannot expect complete agreement on all aspects of transport. However, it would be extremely valuable for the ecosystem if datacenter operators knew whether to implement host-based load balancing, or require specific primitives from switch vendors. This paper is about providing a technical foundation for such a decision.

Our analysis yields several key findings regarding the landscape of load balancing for AI training:
\begin{itemize}
    \item \textbf{Performance hierarchy:}  Among existing contenders, packet-spraying approaches consistently dominate alternative strategies. Between in-switch \vs host-based packet spraying, we find that their performance differs primarily under failure: host-based approaches generally dominate, although both alternatives perform similarly if the underlying routing convergence time is very small.
    \item \textbf{Queueing delay:}
    We show that none of the commonly deployed load-balancing techniques---including packet spraying---are optimal in the sense of maintaining a bounded queue size while operating at maximum network utilization.
    \item \textbf{Optimal scheduling:} We demonstrate that a \textit{destination-based rotation (\DR)} scheduling discipline achieves \(O(1)\) queue bounds. %
    \item \textbf{Practical realization:} We present \name, a novel switch-based realization of \DR scheduling. We show that \name is both practically deployable and theoretically optimal.
 
    \item \textbf{Sensitivity analysis:}  Beyond direct performance comparisons, our results expose how key parameters---buffer size, message size, collective type, \etc---impact performance without fundamentally altering the above core findings. 
\end{itemize}

The remainder of this paper is organized as follows: 
\S\S\ref{sec:background}--\ref{sec:space} establish AI workload requirements and the load-balancing design space; 
\S\ref{sec:approach} details our methodology and 
\S\ref{sec:lb} compares existing load-balancing schemes; 
\S\ref{sec:optimality} analyzes the optimality of different load-balancing schemes and, in \S\ref{sec:algo}, we describe \name; 
\S\ref{sec:sensitivity} presents a detailed sensitivity analysis of the various load-balancing approaches that we evaluate. 

\textit{Generative AI was used to refine text and generate code for this work. This work does not raise any ethical concerns. }

\section{Background: AI Workloads}
\label{sec:background}
AI workloads have specific traffic patterns and performance metrics that are quite different from traditional datacenter workloads.
More specifically, AI workloads involve rounds of computation interspersed with rounds of communication, where the communication is one of a few basic collectives, such as AllReduce, AllGather, and AlltoAll (see \cite{ncclcollectives, parallelism} for a review). 
Thus, the communication workloads can be well characterized by the traffic generated by these collectives, and the network-related performance metric is the collective completion time (CCT) since stragglers will slow the overall training progress.
This results in workloads that are:
\begin{itemize}
    \item \textbf{Uniform:}  The nodes participating in a collective are typically using the same communication library (\eg \cite{nccl}) and the same lower level networking support (\ie transport protocol and NICs). The collectives are commonly symmetrical, in that each participating node has the same amount of data to send, and generates traffic with the same traffic characteristics, with uniform packet size.
    
    \item \textbf{Synchronous:} All nodes participating in the collective begin at roughly the same time. This is particularly relevant to load balancing, because it synchronizes the traffic \cite{ethereal}, which is both good for load balancing (no need to adapt to newly arriving flows while a collective is in progress) and bad (creating possible incast congestion).
\end{itemize}

\mypar{Mixture-of-Experts (MoE)} 
Since we focus on AI \emph{training} workloads, we assume that traffic is well approximated by uniform  workloads (specifically, AlltoAll collectives), even for MoE architectures. This assumption is justified because, with a large enough number of tokens, experts will be relatively uniformly load-balanced \cite{wang2024auxiliarylossfreeloadbalancingstrategy} and/or experts will be split among GPUs \cite{deepspeed-moe}. Existing work in AI network-level load balancing makes a similar assumption~\cite{bonatoreps, ethereal}.

We note that AI workloads have two characteristics that accentuate the limitations of the single-path paradigm. First, the vast bulk of the AI traffic is in long-lived flows that do not have idle periods, so techniques
that improve load-balancing by rerouting flows during idle times (e.g., PLB, flowlet switching) may be less effective. 
Second, the performance goal with these workloads is to minimize the completion of \emph{all} flows in a communication collective collective; i.e., the CCT, which is quite different from the traditional focus on per-flow metrics (e.g., flow-completion time and per-flow fairness).

AI workloads above a certain scale are often deployed in isolation \cite{meta, alibaba, openai-ms-blog}. This is the case we assume in our paper, because it is in these large-scale isolated deployments that achieving high performance (\ie low CCTs) is so critical.

\section{Load Balancing Options}
\label{sec:space}

We start this section by reviewing the overall design space for load balancing (LB), and then present the specific designs that our evaluation focuses on. 

\subsection{Design Space}
The LB design space has been outlined (at least in part) in several papers \cite{clove, letitflowlet, conga}, so here we provide a fairly brief summary. At a high level, the main dimensions of the design space are:
(1) the \textbf{granularity} of traffic that is placed on different paths, which can vary from a packet (\eg \cite{ndp, eqds, dcpim}) to entire flows (\eg \cite{ecmprfc}), with various intermediate granularities such as ``subflows'' (\ie parallel streams whose union reconstitutes the original connection \cite{mptcp}) or ``flowlets'' (small sets of contiguous packets~\cite{kandulaflowlet,plb}); (2) the \textbf{location} at which LB decisions are made can be centered at the host (\eg \cite{bonatoreps, plb}) or in switches  (\eg \cite{spectrumx, pfabric}); (3) how LB \textbf{adapts} to varying network load, with some schemes adapting to local state at a host or switch (\eg \cite{plb, clove, localflow}), others to global state (\eg \cite{conga, pfabric}), and yet others are not adaptive at all (\ie use no state about load, \eg \cite{ecmprfc, ndp, rps}).\footnote{We note that all designs use state about network connectivity to determine the current set of equal cost paths, so the adaptivity we consider here is whether/how different schemes react to state about the \emph{load} on the network.}

Of course, not all combinations of points along these dimensions lead to sensible solutions. Nonetheless, this is clearly a large design space and there is a vast literature exploring it. Table~\ref{tab:designspace} references just a sample of such work. For tractability, we focus our study on a key set of design choices that are being evaluated or deployed in current practice. These are indicated with a $\checkmark$ symbol in Table~\ref{tab:designspace} and, as we can see, span a range of our design dimensions. We call these our \textit{leading contenders}  and describe them in more detail next.

\begin{table*}[]
    \caption{A survey of existing load balancing approaches based on their considered state, granularity, and implementation location. We consider the design points marked with \checkmark, and introduce the one with \textdagger. We note that some works allow splitting flows in time/packets in different amounts and with different strategies, thus we use a wide definition of ``flowlet'' that includes any splitting in these dimensions. 
    We do not include \emph{routing state} in our definition of global state for switches (only congestion/traffic information) since it is often locally-available in the switch despite being a function of global information.}
    \vspace{-1em}
    \label{tab:designspace}
    \centering
    {
    \footnotesize
    \begin{tabular}{c c l l l l l l l l} 
    \toprule
         & \textbf{Location} & \multicolumn{4}{c}{\textit{Host}} & \multicolumn{4}{c}{\textit{Switch}} \\
         
         \cmidrule(r){3-6} \cmidrule(l){7-10} 
         
        & \textbf{Granularity} &  \multicolumn{1}{c}{\textit{Packet}} & \multicolumn{1}{c}{\textit{Flowlet}} & \multicolumn{1}{c}{\textit{Subflow}} & \multicolumn{1}{c}{\textit{Flow}} & \multicolumn{1}{c}{ \textit{Packet}} & \multicolumn{1}{c}{\textit{Flowlet}} & \multicolumn{1}{c}{\textit{Subflow}} & \multicolumn{1}{c}{\textit{Flow}} \\
        \midrule
        
        \multirow{3}{*}{\textbf{State}} & \textit{None} & \checkmark \cite{ndp,eqds} & $\times$ \cite{presto,letitflowlet} & \checkmark \cite{mptcp} & $\times$ & \checkmark \cite{pfabric,rps,dcpim} & $\times$ & $\times$ & \checkmark \cite{ecmprfc} \\
        
        & \textit{Local} & \checkmark \cite{bonatoreps,strack} & \checkmark \cite{plb, flowbender, hermes, clove} & $\times$ & $\times$ & \checkmark \textdagger \cite{spectrumx,detail} & $\times$ \cite{kandulaflowlet, conweave} & $\times$  \cite{localflow} & $\times$ \cite{broadcomtom} \\
        
        & \textit{Global}  & \checkmark\cite{fastpass, drb}  & $\times$& $\times$ \cite{ethereal} & $\times$ \cite{mahout}& $\times$ \cite{bytedance-glb} & $\times$\cite{conga,hula,expeditus} & $\times$ & $\times$\cite{hedera, microte}\\
         \bottomrule
    \end{tabular}
    }
\end{table*}

\subsection{Leading Contenders}
We focus on the following seven schemes and use reference implementations (particularly from \htsim~\cite{csghtsim}) when possible.

\noindent \textbf{(1) Per-flow LB:} Here, packets are hashed based on their five-tuple + flow label and placed on one path in the network. This approach captures the behavior of ECMP and, per the dimensions discussed above, represents LB at the granularity of flows, that is non-adaptive (\ie stateless), and only requires simple hashing at switches. 

\noindent \textbf{(2) Per-subflow LB:} This scheme operates as above, except that a host now divides a flow into $n$ parallel subflows, each with a unique flow label. This approach captures protocols such as MPTCP~\cite{mptcp} and represents LB at the granularity of subflows, that is non-adaptive (stateless), and where parallelization into subflows is controlled by hosts while switches only implement simple hashing (as above).

\noindent \textbf{(3) Per-flowlet LB: } Modeled after PLB~\cite{plb}, in this scheme, the host \emph{may} change the flow label if it detects congestion above a certain threshold. Specifically, in our implementation, a host will change the flow label at most every $\alpha$ packets if over $\beta$\% of recent packets carried ECN marks with $\alpha$ and $\beta$ picked based on the test  environment (\S\ref{sec:lb}).\footnote{This rule differs from PLB in one aspect: in PLB, a host only changes the flow label at RPC boundaries; since we don't have RPC traces, we instead allow the host to change a flow label at most every $\alpha$ packets.} 

\noindent \textbf{(4) Host Per-Packet LB:} In this scheme, the host simply changes the flow label on every packet. This represents load balancing at the granularity of individual packets, that is implemented at hosts, and is non-adaptive. %

\noindent  \textbf{(5) Switch Per-Packet LB:}  In this scheme, the switch implements simple round-robin scheduling of packets across the set of next-hop ports (\ie ports associated with the available equal-cost paths to a  destination), where the round-robin order is periodically reset to avoid synchronization effects. This implementation is borrowed from the \htsim simulator~\cite{csghtsim}.

\noindent \textbf{(6) Adaptive Host Per-Packet LB:} Based on the REPS protocol~\cite{bonatoreps}, the endhost in this scheme sets flow labels randomly at first, then 
upon receiving ACKs, discards flow labels that were ECN marked and reuses the uncongested flow labels. 

\noindent \textbf{(7) Adaptive Switch Per-Packet LB:} Here, the switch places the packet on the next-hop port with the shortest queue, where the shortest queue is determined by first quantizing the queue length into a few bins determined by some thresholds (\eg 0-5\%, 5-10\%, 10-20\%, and >20\%) and picking randomly among the options in the smallest queue bin  \cite{dragonfly-ar, nvidia_AR_whitepaper, csghtsim}.

\section{Approach: Decoupling Components}
\label{sec:approach}

We aim to isolate the merits of LB from the effects of two other components of the network stack: \textit{congestion control} (how hosts adapt their rate to avoid congestion), and \textit{loss recovery} (includes both loss detection and packet retransmission). We now explain our methodology for doing so.

\T{Congestion control:} Our approach is based on the observation that a perfect LB scheme would make the entire capacity of the network available to hosts. Thus, if we were to assume: (i)~perfect LB, (ii)~uniform collective workloads, and (iii)~no failures, then, the job of a CCA would be  quite simple: every host transmits at the same fixed rate, which is calculated by dividing the available capacity equally across all hosts. In such a scenario, any drops or delays can be attributed to imperfect LB rather than an imperfect CCA (\ie imperfections in host sending rates). 

Hence, to isolate the impact of LB, we simulate an idealized fixed-rate CCA in which each host simply sends at a rate that is set to its share of the available capacity of the physical topology. Specifically, for a full bisection bandwidth  network without failures, each host sends at its line rate. 
In failure scenarios, there may not be sufficient capacity for the workload to proceed at line rate. 
Thus, we calculate the maximum possible sending rate $\rhomax$ at which all flows could send while load-balancing equally across all valid shortest paths, and all hosts send traffic at a rate of  $\rhomax$. The process of determining $\rhomax$ is presented in Appendix~\ref{app:failure}.

Finally, in addition to the above ideal CCA, we also later consider \textbf{practical CCAs}. For this, we would like to compare our LB options using Google's Swift~\cite{swift}, which combines low delays and high throughput. However, Swift is designed for single paths and experiences throughput collapse when handling spraying. Therefore, we use MSwift~\cite{mswift}, a recent and publicly-available CCA that extends Swift to operate under spraying.

\T{Loss recovery:} 
Packet loss is inevitable due to link failures. And, some (imperfect) LB may cause sufficient queueing to cause congestive loss. The job of loss recovery 
is to detect which packets have been lost and schedule their retransmissions.  Given the potential for packet-level reordering under fine-grained LB, distinguishing loss from reordering can be challenging.

Following the same strategy as above, we consider loss recovery approaches representing ideal and realistic options. To model an ideal loss recovery solution, we simulate a mechanism based on \textit{rateless erasure coding}~\cite{raptor-theory, raptorq-rfc, ltcodes}  which ignores losses and just transmits packets until the file is complete. In this case, the sender encodes the data before sending and creates some number of recovery symbols that can be used to recover lost data. The receiver can decode the entire messages with high probability as long as a sufficient number of symbols have been received. Therefore, such a protocol comes with a constant overhead (the required recovery symbols), but does not need to manage specific packets/sequence numbers (all symbols are equally necessary and can be lost). %
To evaluate an LB scheme without the impacts of loss recovery, we simulate ideal erasure coding with no overhead. For this evaluation, we unrealistically assume that no additional decoding symbols are necessary to receive the message (\ie only a number of bytes equal to the message size must be delivered to decode). Thus, the sender simply sends packets until the receiver has ACKed the full message and no other mechanisms are needed to recover from loss.\footnote{Given the delay between when the sender transmits the final \emph{necessary} packet and when it receives the corresponding ACK, some additional data may be transmitted. To avoid extra transmissions that may cause unnecessary queueing, our implementation opportunistically stops in the final BDP and only sends more if necessary.}

For our realistic options, we first consider a \textit{SACK}-based protocol since this is used in many recent protocols to tolerate out-of-order delivery~\cite{falcon, rocesack, sack, uec-spec}.
In this case, packets have a bitmap field that can tell the sender exactly which packets have not been received. The sender retransmits all missing packets when the gap between the lowest and highest acknowledged sequence numbers crosses a configured threshold.  Importantly, these retransmission triggers do not change the pacing rate, only which packets are sent.

Our second realistic option adds the use of \textit{MSwift} as a CCA so that, unlike the previous solution, we now capture the effect of varying sender rates. MSwift also uses a SACK-based loss recovery mechanism. %

\section{Results: Load Balancing Schemes}
\label{sec:lb}
As discussed before, we focus our initial evaluation on 7 leading contender designs with key parameters as shown in Table~\ref{tab:contenders}. We use the default parameters for existing schemes where possible and otherwise swept for well-performing configurations; different configurations may produce different results.

\mypar{Simulator} We compare different load balancing schemes with simulations on \htsim~\cite{csghtsim}, which we chose due to its ability to scale to large datacenter networks and its existing implementation of many design points in this space. 
Specifically, unless stated otherwise, we simulate a 128-node 3-tier fat-tree network with $800\unit{Gbps}$ links, $0.5\us$ link latency, $800\unit{KB}$ per-port
buffers,\footnote{This value roughly follows the per-port per-Gbps buffer size of modern switches~\cite{broadcomtom,spectrumx}, is above a BDP, and resulted in no loss in most evaluated cases.} 
RoCE-v2-sized packets with $4,096\unit{B}$ payloads and $62\unit{B}$ headers, $64\unit{B}$ ACKs, and $20\unit{B}$ gaps between frames ($12\unit{B}$ inter-frame gap and $8\unit{B}$ preamble and start frame delimiter).
We perform a sensitivity analysis for many of these parameters in \S\ref{sec:sensitivity}.
We run all considered scenarios 10 times to determine the level of variability in the measured performance.

\begin{table}
\caption{Evaluated LB leading contenders and the applicable configuration. Default configurations were generally found by taking standard thresholds or experimenting with values and taking the best-performing.
}
\vspace{-1em}
\label{tab:contenders}
\centering
{
\footnotesize
    \begin{tabular}{l l  >{\raggedright\arraybackslash}p{3cm}}
    \toprule
    \textbf{Approach} & \textbf{Modeled After} & \textbf{Simulated Configuration} \\ 
    \midrule
    \flow & ECMP & - \\ \arrayrulecolor{black!20}\hline
    \subflow & MPTCP~\cite{mptcp} & \# subflows = 4 \\ \arrayrulecolor{black!20}\hline
    \flowlet & PLB \cite{plb} & ECN thres.: 50\%, label change thres.: 40\% \\ \arrayrulecolor{black!20}\hline
    \hpkt & OPS \cite{rps} & - \\ \arrayrulecolor{black!20}\hline
    \spkt & Round-robin~\cite{csghtsim} & Permute every 5 wraparounds \\ \arrayrulecolor{black!20}\hline
    \hpktar & REPS \cite{bonatoreps} & ECN thres.: 10\% \\ \arrayrulecolor{black!20}\hline
    \spktar & Spectrum-X \cite{spectrumx} & Quanta: 5\%, 10\%, 20\% \\
    \arrayrulecolor{black}\bottomrule
    \end{tabular}
}
\end{table}

\mypar{Workloads} 
In our evaluation of load balancing schemes, we consider an all-to-all (ATA) traffic matrix and a random permutation matrix. In the permutation traffic, each node is sending traffic to exactly one other node, and receiving traffic from exactly one other node. 
Rather than modeling specific implementations of every collective, we select these two traffic matrices since they are representative of many collective algorithms. Permutation matrices are often used to implement collectives such as AllGather and AllReduce \cite{ring-mpich, nccl-ring, sccl, meta} and can be used to iteratively perform an AlltoAll ($n-1$ permutation matrices iteratively). Similarly, an all-to-all traffic matrix may implement AllGathers or a one-shot AllReduce, as well as an AlltoAll.
Notably, in most training jobs, multiple forms of parallelism will be used and thus no one collective will span the whole network. We explore this more in \S\ref{sec:sensitivity} and for now make the simplifying assumption of one collective at a time across all nodes. We expect smaller slices of the network (\eg a pod) to exhibit similar behavior.

\mypar{Metric} For any given LB scheme, we report the percentage increase in completion time relative to the lowest possible completion time given the host-ToR link rate and propagation delay of the network (\ie with no loss or queueing delay). For example, in the all-to-all traffic matrix, the minimum possible completion time in our setup is $\sim1.3$ms. Thus an increase of (say) $10\%$ in our normalized completion time metric is a difference of $130$us in which a GPU could perform billions of FLOPs \cite{h100}. Such differences could have a noticeable impact on training time as the communication time compounds over iterations.
Our all-to-all CCT lower bound is computed as the simple sum of propagation and host transmission delays. The permutation CCT lower bound is computed similarly but additionally accounts for the symmetric dynamics of data and ACK packets, as detailed in \cref{sec:lowerbound}.

\subsection{Performance without failures} \label{sec:default}

\begin{figure}
    \centering
    \begin{subfigure}[b]{0.49\columnwidth}
        \includegraphics[width=\columnwidth]{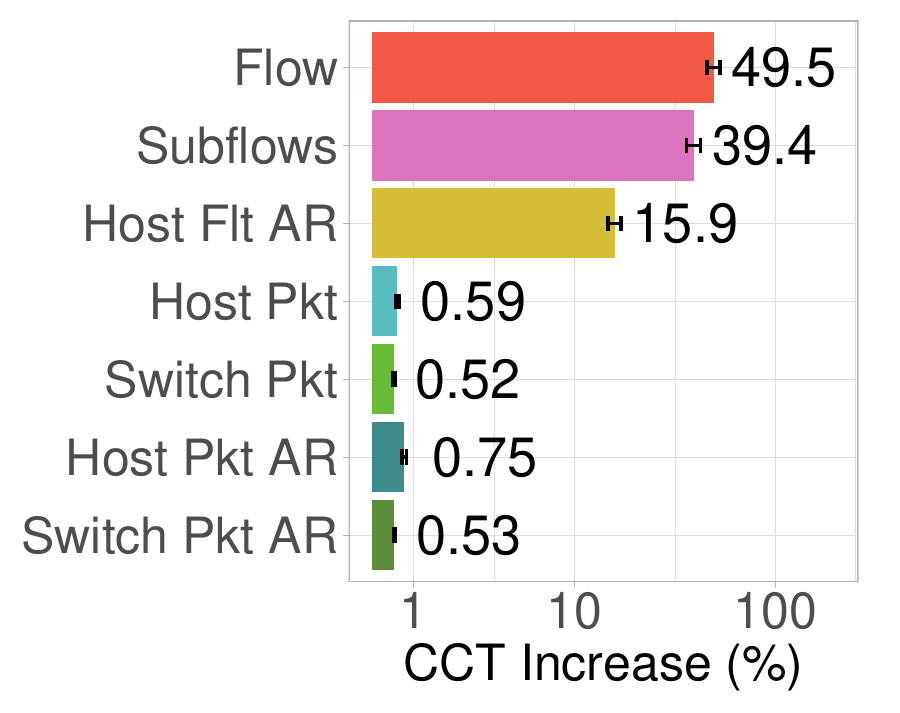}
        \caption{All-to-All}
        \label{subfig:ata-baseline}
    \end{subfigure}%
    \begin{subfigure}[b]{0.49\columnwidth}
        \includegraphics[width=\columnwidth]{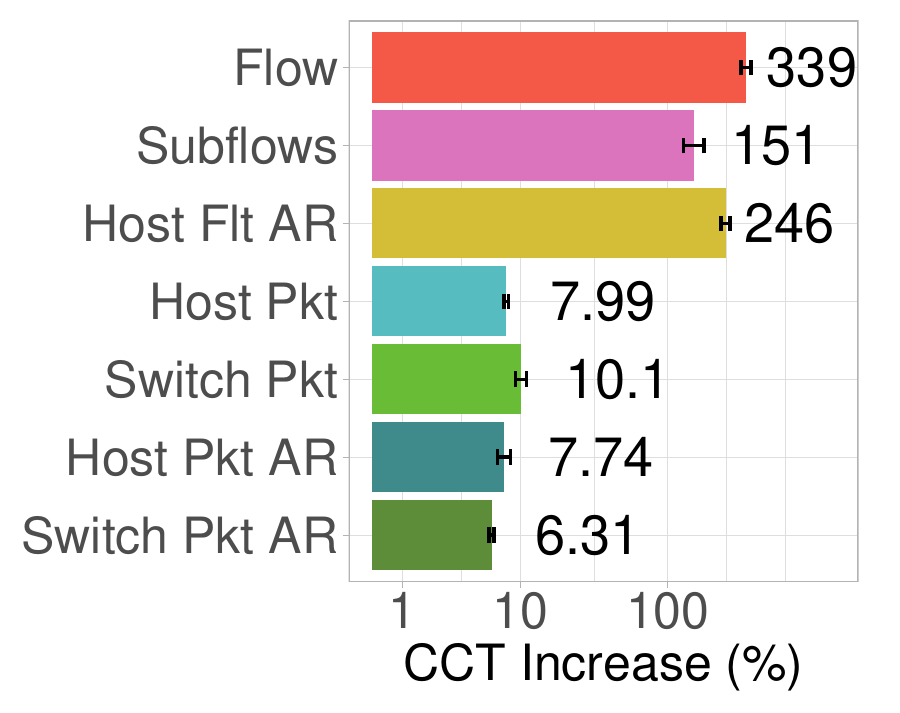}
        \caption{Permutation}
        \label{subfig:perm-baseline}
    \end{subfigure}
    \caption{Completion times (increase over ideal) for different load balancing schemes with a perfect loss recovery protocol. Adaptive techniques are denoted with ``AR'' (adaptive routing). 
    }
    \label{fig:baseline}
\end{figure}

Figure~\ref{fig:baseline} shows the performance of our leading contenders in our  default scenario in the absence of failures.
As one might expect, we see that load balancing at finer granularity offers significant benefits with per-packet LB outperforming flowlet-, subflows-, or flow-based LB. 
We see little difference across the various packet spraying approaches which, again, is perhaps to be expected in the absence of failure.%

Comparing performance across collectives (\cref{fig:baseline}(a) \vs (b)), we see that the normalized differences are larger for the permutation matrix since the minimum completion time is much smaller with fewer flows competing for bandwidth. 
As we explore more in \S\ref{sec:sensitivity}, the relative gap decreases with larger message sizes and increases with larger networks.
Notably, in the ATA, packet-based schemes achieve within 1\% of the lower bound. 
We also see that adding subflows in hopes of increasing the entropy helps more in the permutation matrix as the ATA already has many flows.
Finally, performance of \flowlet varies greatly across the two collectives: it achieves within 16\% of optimal in ATA, but does worse than \subflow in the permutation matrix. This is, in part, because in the permutation matrix, most of the flow is sent in the first BDP and thus the feedback to move paths comes quite late.

\subsection{Performance with failures}\label{sec:fail}
We now look at whether our previous findings hold in the face of link failures. Failure cases pose a particular challenge to the LB schemes as the network is no longer symmetric (\ie not all links should carry the same amount of traffic).

In general, we find that the gap between packet \vs coarser (flow, subflows, flowlet) granularities of LB only increases under failures and hence we focus our evaluation on the four different forms of packet spraying designs.
In particular, we want to understand the differences between host-  \vs switch-based spraying approaches: if they continue to perform similarly, operators might prefer host-based approaches since that allows them to be vendor agnostic. 

\begin{figure}
    \centering
    \includegraphics[width=0.7\linewidth]{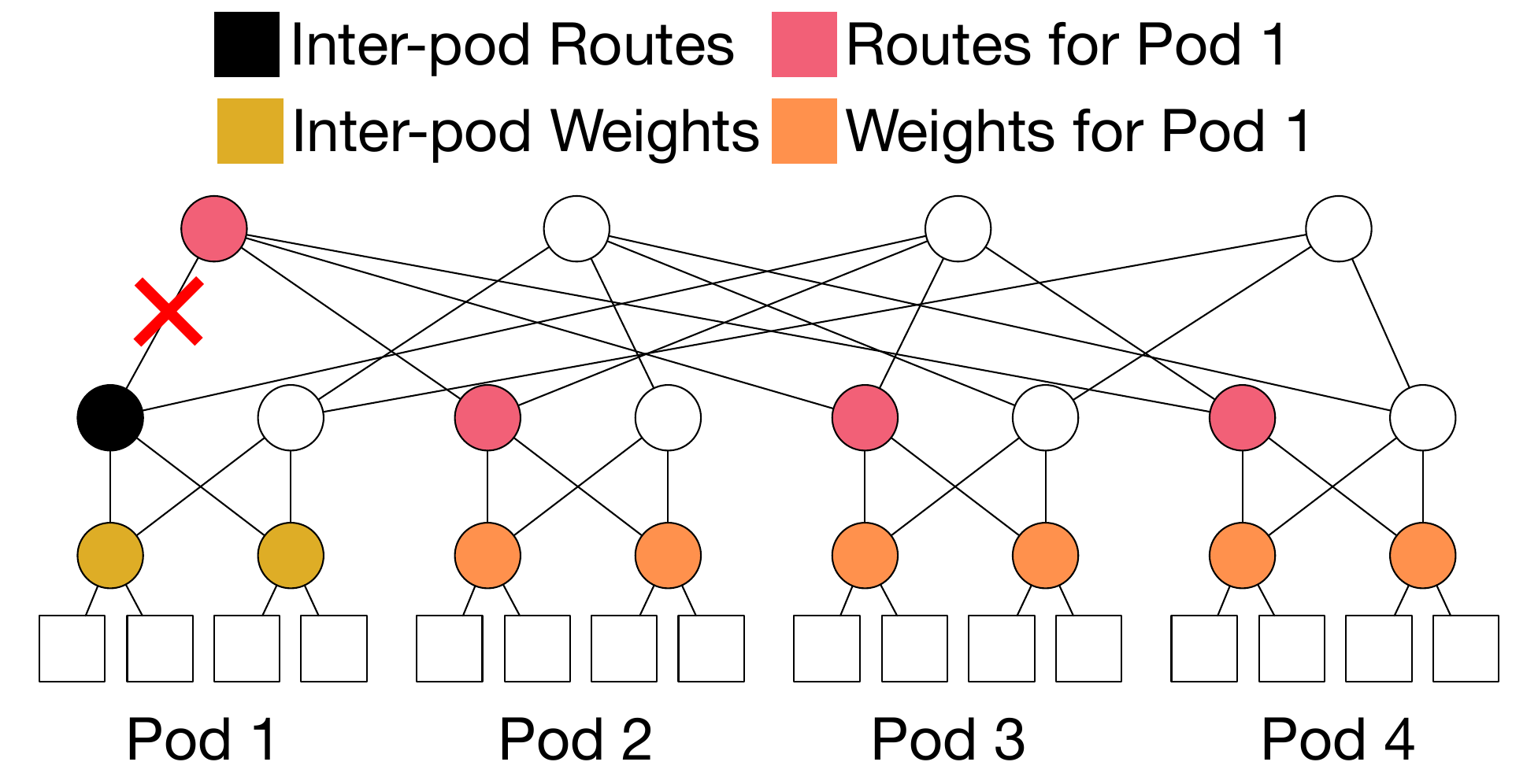}
    \caption{Switches that must update routing and/or load balancing state based on a link failure. The pod with the failure must update state for all routes, and many nodes in other pods must update routes and weights for traffic destined to Pod 1.
    }
    \label{fig:failure-diagram}
    \vspace{-1em}
\end{figure}

\mypar{Failure models} We consider randomized link failures in which 
we fail each edge-aggregation and aggregation-core link with a given probability. 
We initially set the probability to 1\% \cite{gill2011understanding} and explore more values later in the section.

\mypar{Modeling routing convergence} 
While routing and load-balancing can be viewed as orthogonal, the performance of the routing layer impacts the performance of load balancing under failure. I.e., performance under failure will depend on the delay from when a failure occurs to when switches update their routes/weights to reflect the failure. 

In a datacenter, when a link $l$ at a switch $S$ fails, recovery involves two key steps. 
First, $S$ must detect the failure and remove $l$ as a next-hop option from its relevant port groups. This in itself can take some time (estimates range from 10s of microseconds \cite{bytedance-glb} to 10s of milliseconds \cite{ethereal, bonatoreps} depending on the underlying system) and during this period, packets transmitted over the failed link are silently dropped.

In addition, all \emph{other} switches in the network must be notified of the failure so that they can update their routes and weighted ECMP (W-ECMP) configuration accordingly.  An example of this process is shown in \cref{fig:failure-diagram} in which route updates are propagated to all aggregation switches connected to the same core router and all ToRs update how they weight each of their aggregation switches for Pod 1. Datacenters commonly run a routing protocol (\eg BGP~\cite{bgp-in-dcs, bgp-in-dcs-rfc}) whose convergence time will determine how quickly switches update their routes and weights. 
As above, a wide range of cross-switch convergence times have been reported from 100s of microseconds~\cite{bytedance-glb} to 100s of milliseconds~\cite{bgp-in-dcs}.

Given the range of reported convergence times, we are interested in how this impacts LB performance.
For example, REPS~\cite{bonatoreps} motivates its host-based approach with potentially long convergence time.
Accordingly, if the convergence time could be decreased with other in-switch systems \cite{bytedance-glb}, would host-based AR still be important for performance under failure?
Thus, in our simulations, we introduce a parameter $G$ to denote this global convergence time. We make the simplifying assumption that this convergence time is the same across all switches, including (pessimistically) the local switch $S$. 
We measure the impact on CCT as we vary $G$, under our randomized link failure model.

\mypar{\underline{$\boldsymbol{G=\infty}$}} 
\begin{figure}
    \centering
    \begin{subfigure}[b]{0.48\columnwidth}
        \includegraphics[width=\columnwidth]{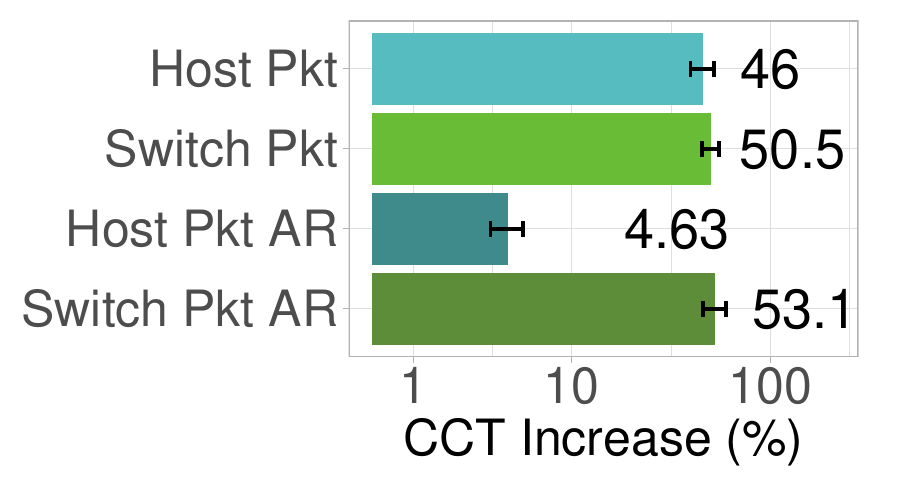}
        \caption{All-to-All}
        \label{subfig:ata-fail-rand}
    \end{subfigure}%
    \begin{subfigure}[b]{0.48\columnwidth}
        \includegraphics[width=\columnwidth]{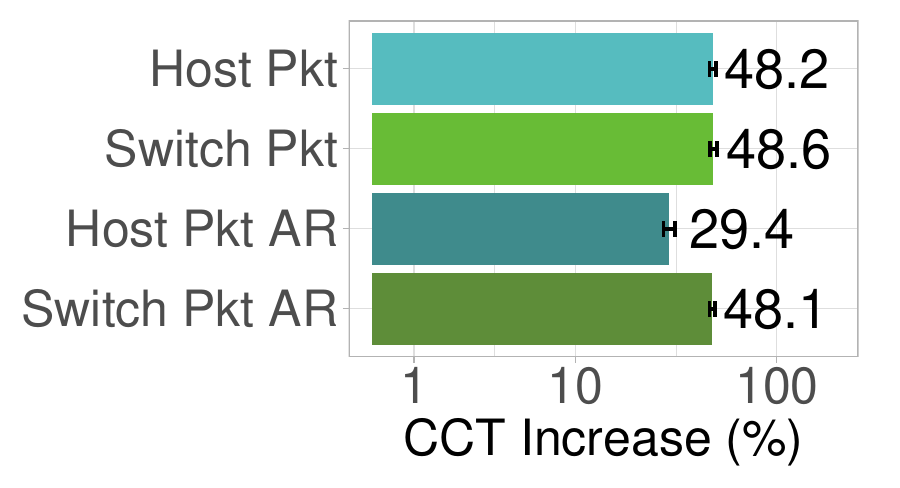}
        \caption{Permutation}
        \label{subfig:perm-fail-rand}
    \end{subfigure}
    \caption{CCT increase for randomized failures with a per-link failure rate of 1\%, with $\boldsymbol{G=\infty}$ (\ie convergence time $>$ CCT).}
    \label{fig:failure-random-g0}
\end{figure}
This represents scenarios in which routing state remains broken during the entire collective runtime. This scenario is relevant to practice, given reports of convergence time that can extend into several 100s of milliseconds, on par with expected CCTs.

Our results in \cref{fig:failure-random-g0} now show a clear winner: \hpktar. This is 
because, on the timescale of a small number of RTTs, \hpktar is able to detect the failure (via timeouts on a particular path label) and avoid that path in future transmissions. 
Notably, \spktar, even though it is adaptive, does not see this benefit because the queue length on the failed link may be short and thus continues to be chosen. Intuitively, this performance difference arises because \hpktar enjoys an end-to-end view of a path's performance (and can react accordingly) while \spktar is limited to its local view of switch state. Further, with \spktar, even if the switch with the failed link had reacted quickly to avoid the failed link locally, it would still receive traffic from other switches as if the link was up, thus causing inevitable overload. 

Finally, we note that the performance degradation due to slow convergence can be substantial, even in the best case of adaptive host-based solutions; \eg 20\% higher than the baseline for the permutation matrix.

\mypar{\underline{Varying $G$}}
\begin{figure}
    \centering
    \begin{subfigure}[]{0.75\columnwidth}
        \includegraphics[width=\columnwidth]{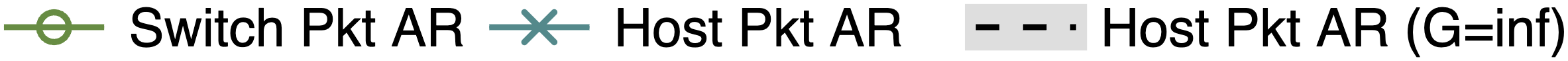}
    \end{subfigure}
    \begin{subfigure}[b]{0.48\columnwidth}
        \includegraphics[width=\columnwidth]{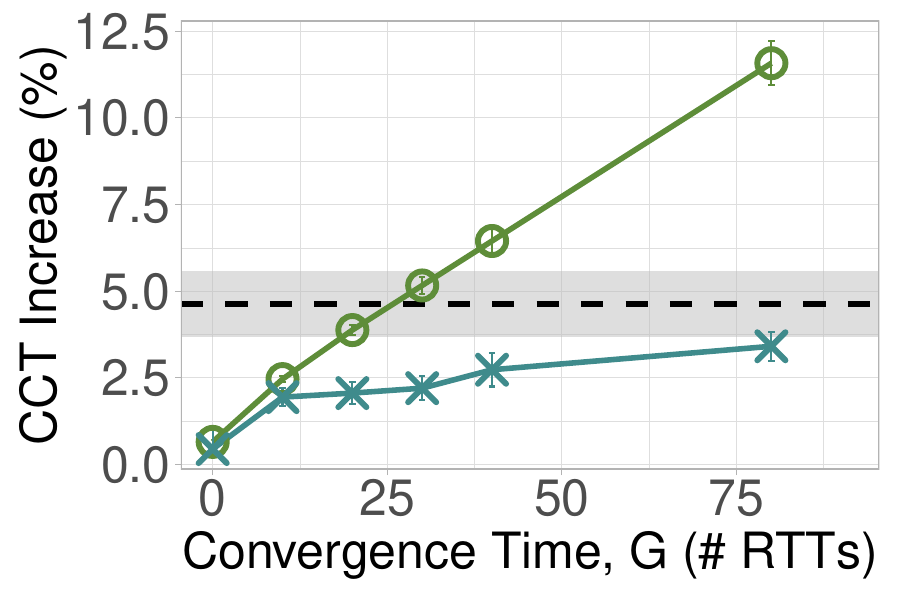}
        \caption{All-to-All}
        \label{subfig:ata-fail-time-rand}
    \end{subfigure}%
    \hspace{0.02\columnwidth}
    \begin{subfigure}[b]{0.48\columnwidth}
        \includegraphics[width=\columnwidth]{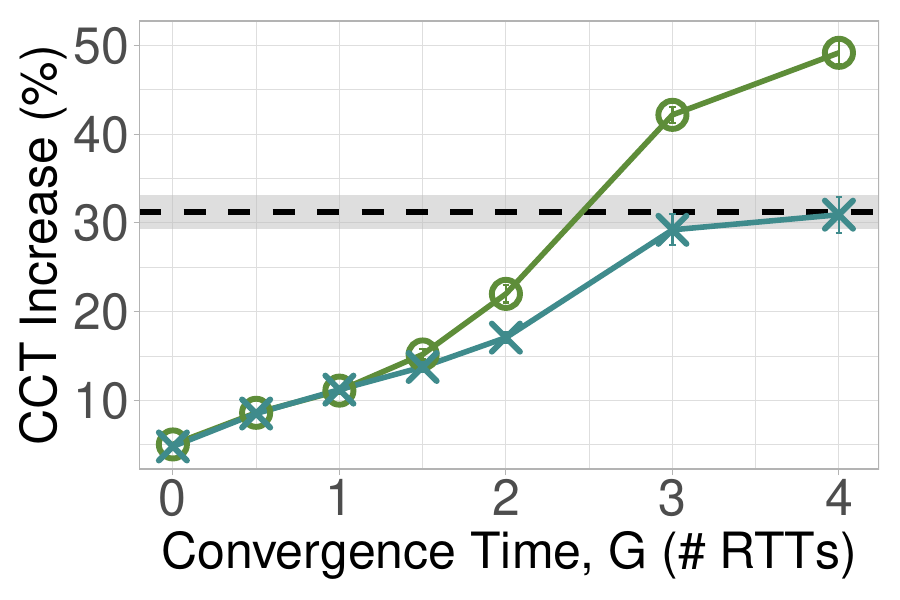}
        \caption{Permutation}
        \label{subfig:perm-failtime-rand}
    \end{subfigure}
    \caption{CCT increase over best possible under random link failures (1\% fail rate) for host adaptive (dashed and shaded) compared to an in-network rebalancing approach with different delays (x-axis). The CCT lower bound accounts for the failure (uses $\rhomax$).
    }
    \label{fig:failure-time-rand}
    \vspace{-1em}
\end{figure}
Figure~\ref{fig:failure-time-rand} shows the performance for different convergence times $G$, focusing only on the adaptive schemes (\spktar and \hpktar) for simplicity. We set $G$ to be a (varying) multiple of the minimum network RTT. Hence, as we increase $G$, a growing fraction of the collective runtime overlaps the network's convergence period. For calibration, the minimum CCT (with no failure) for the ATA is about 220 RTTs, while the permutation matrix is between 2-3 RTTs. 
We include the performance of \hpktar \emph{with no convergence} ($G = \infty$) for reference.
In general, we see that at very low $G$, switch-based and host-based approaches perform very similarly, 
but host-based solutions perform best  as G grows. 
While routing convergence may re-hash flow labels previously categorized as 'good' or 'bad' by the host LB algorithm onto new paths, performance is maintained because W-ECMP weights now correctly distribute traffic across all available capacity; the only residual effect is that the host-based approach continues to exclude the specific label it identified with the failure.
With the ATA, the purely switch-based approach eventually performs worse than Host AR with $G=\infty$.
Meanwhile, because the CCT for the permutation matrix is smaller, we require very  fast convergence times (1 RTT) to limit performance degradation. For large message sizes on a permutation matrix, the acceptable convergence time is longer, similar to the ATA.

\mypar{\underline{$\boldsymbol{G=0}$}}
As shown in \Cref{fig:failure-time-rand}, when $G=0$ (\ie routes are converged before the collective even begins) both approaches perform  similarly. Thus, in-switch and end-host based packet spraying are similar in the no-failure case and under failures if the routes have converged.

\begin{figure}
    \centering
    \begin{subfigure}[]{0.55\linewidth}
        \includegraphics[width=\linewidth]{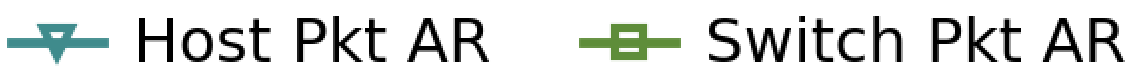}
    \end{subfigure}  
    \par    
    \begin{subfigure}[t]{0.55\columnwidth}
        \centering
        \includegraphics[width=\linewidth]{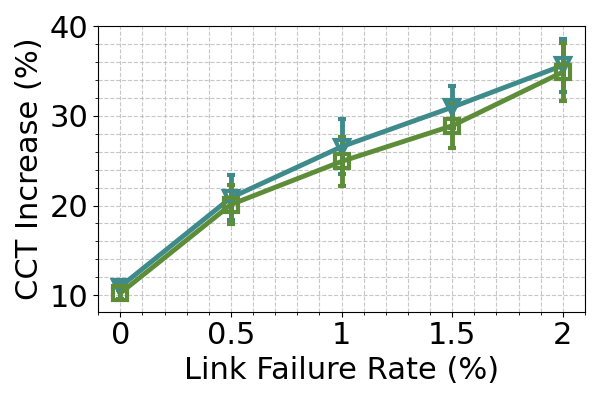}
 
    \end{subfigure}
    \caption{Impact of randomized link failure rates with $G=0$. CCT is normalized to best-case \emph{without} failures.}
    \label{fig:fail}
\end{figure}

\mypar{\underline{Varying failure rate; $\boldsymbol{G=0}$}} 
Finally, \cref{fig:fail} plots performance under increasing link failure rates when routing state has converged ($G=0$). 
We see that, as expected, performance degrades with increasing failure rate but the performance gap (or lack thereof) across different schemes remains unchanged.

\subsection{Takeaways}
Overall, our comparison of leading LB approaches yielded a few key insights. 
First, per-packet techniques dominate coarse-grained LB schemes, and, in failure-free scenarios, most existing per-packet techniques achieve similar performance.
While per-packet approaches near optimal performance for ATA, all versions  incur a larger overhead under our permutation traffic matrix. 
Second, we find that the speed of route and (W-ECMP) weight convergence is a critical performance determinant under failure. 
Under very fast convergence times (\eg $<$ 10 RTTs for our ATA and $<$ 2 RTTs for permutation), \spktar and \hpktar perform similarly but, otherwise, \hpktar outperforms. Finally, \spktar and \hpktar perform similarly once all routes have converged.
More generally, we see that rapid convergence significantly improves performance in all cases and hence is an important metric for switch vendors to optimize.

\section{Optimality in Load Balancing}
\label{sec:optimality}
Our results in the previous section revealed that even the best-performing packet spraying approaches can suffer from non-trivial performance overheads, suggesting that there is room for further optimization. In this section, we present theoretical analysis that gives us a more rigorous understanding of optimality in load balancing approaches, and leads us to propose \sysname, a new and provably optimal approach, in \S\ref{sec:algo}.

\subsection{Theoretical Results}
\label{sec:theory}

We define optimality in terms of the average queue length that results from a load balancing scheme when running the network at maximum utilization. Specifically, for a message size $m$, we focus on $\q$, the average queue size across \emph{all} switch queues over the time it takes for \emph{all} hosts to finish sending their messages. With this definition, an optimal load balancing algorithm is one that achieves an $O(1)$ queue size, independent of $m$.

We focus on queue size for a few different reasons. First, queue size is strongly correlated to CCT: queue size is the dominant variable component of the CCT given a fixed propagation delay  and our simplifying assumptions of  uniform traffic, ideal sending rates, and zero-overhead loss recovery. We further validate this in simulation in \S\ref{sec:theory-sims}. In addition, achieving an $O(1)$ queue is a valuable goal in itself since it ensures predictable performance, 
simplifying downstream tasks such as congestion or loss detection, \etc 
Nonetheless, we note that our focus on queue size is largely to build intuition in designing load balancing schemes, and we return later to comparing designs in terms of their CCT.  %

We present the details of our theoretical analysis in \cref{sec:sync,sec:sqrt,sec:DRB}  and only provide a high level summary here. Our model considers $n$ hosts implementing a random permutation collective over a three-level $\k$-ary fat tree. We focus on packet-based load balancing and analyze simplified versions of the contender schemes presented in \S\ref{sec:lb}. These simplifications were necessary because our leading contender algorithms include multiple parameters that complicate formal analysis. Specifically, we analyze:

\noindent \textbf{(i) Simple RR:} We model our switch-based packet spraying algorithm using a simple round-robin model in which the switch load-balances packets across a set of next-hop ports in round-robin order. Thus, the simplification relative to our earlier \spkt scheme is that \simplerr does not periodically reset  the round-robin order.

\noindent \textbf{(ii) JSQ:} We model our adaptive switch-based packet spraying algorithm using a Join Shortest Queue (\jsq) model in which the switch selects the next-hop port with the shortest queue, with random tie-breaking between equal-length queues. This simplifies the adaptive switch-based spraying (\spktar) from \S\ref{sec:lb} by removing the quantization of queues. Or, viewed differently, \jsq represents adaptive switch spraying in which the quantization is extremely fine-grained. 

\noindent \textbf{(iii) RSQ:} In Randomized Switch Queueing (\rsq), the switch selects among possible next-hop ports at random. Hence, \rsq captures the opposite extreme of \jsq: it captures our adaptive switch-based spraying scheme but now with extremely coarse-grained  quantization. 

\noindent  \textbf{(iv) Host Pkt}: We model non-adaptive host-based packet spraying with full fidelity: the host randomly sprays packets (by picking random flow IDs for each subsequent packet), hence this scheme is identical to our \hpkt  scheme in \S\ref{sec:lb}. 

With the above, we can prove the following results. 
\begin{theorem}
\label{thm:sync}
\simplerr and~\jsq suffer from a linear growth of the queue size with the message size:
\begin{equation} \label{eq:linear}
    \q = \Theta \para{\m}
\end{equation}
\end{theorem}

\begin{theorem}
\label{thm:sqrt}
\rsq and \hpkt, that are based on random spraying, see the queue grow at least as the square root of the message size: 
\begin{equation} \label{eq:sqrt}
    \q = \Omega \para{\sqrt{\m}}
\end{equation}
\end{theorem}

The above results suggest that none of our leading contender algorithms achieve our goal of $O(1)$ queue size. In particular, the linear growth of \simplerr and \jsq was surprising, especially as the latter is often presented as a near-optimal solution ~\cite{duato650interconnection,nvidia_AR_whitepaper,nvidia_AR_whitepaper2,gomez2007deterministic}. Examination reveals that the linear scaling of \simplerr and \jsq are a result of \emph{synchronization} effects that arise due to the uniform nature of collective workloads; \ie that all senders transmit the same message and packet size, paced at identical rates. This uniformity leads to per-sender ``stickiness'' in which subsequent packets from the same sender are forwarded along the same outgoing port to the same aggregation switch.
Given the nature of a fat tree topology, such packets necessarily arrive at the same aggregation switch in the destination pod, and then ``stick'' to the same path all the way down to the host destination. They then collide with other \textit{sticky} flows. 
We discuss these synchronization effects in detail in \cref{sec:sync} and further validate them in simulation below.

\begin{table}
\caption{Modeled algorithms for analysis. 
}
\vspace{-1em}
\label{tab:models}
\centering
{
\footnotesize
    \begin{tabular}{l l l l}
        \toprule
        \textbf{Approach} & $\q$ & \textbf{Simplification of} & \textbf{Simplification} \\ 
        \midrule
        \simplerr & $\Theta \para{\m}$ & \spkt & No permute \\ \arrayrulecolor{black!20}\hline
        \jsq & $\Theta \para{\m}$ & \spktar & No quantization \\ \arrayrulecolor{black!20}\hline
        \rsq & $\Omega \para{\sqrt{\m}}$ & \spktar & One quantum \\ \arrayrulecolor{black!20}\hline
        \hpkt & $\Omega \para{\sqrt{\m}}$ & - & -\\ \arrayrulecolor{black!20}\hline
        \hdr (DRB) & $\Theta(1)$ & - & - \\ \arrayrulecolor{black!20}\hline
        \name & $\Theta(1)$ & - & - \\ 
        \arrayrulecolor{black}\bottomrule
    \end{tabular}%
}
\end{table}

Table~\ref{tab:models} summarizes our models and findings. 

Given these results on the non-optimality and/or fragility of our leading contenders, we expanded our horizons to the broader scheduling literature in search of approaches that lead to our goal of O(1) queueing. We find inspiration in the DRB scheduler by Cao \textit{et al}~\cite{drb}. The key idea in DRB is that each source host load balances packets in a round-robin fashion on a \emph{per destination} basis. 
We generalize DRB's idea to define a new scheduling discipline that we term \emph{Destination-Based Rotation (DR)}.\footnote{\DR has roots in \textit{load-balanced router} algorithms like UFS, FOFF and Padded Frames~\cite{LBR,thesis,padded-frames}.} We observe that DRB can be viewed as a specific host-based implementation of \DR scheduling, and we later contribute a new switch-based \DR scheme. Since DRB runs at hosts (and to contrast it from the switch-based scheme we propose), we henceforth refer to DRB as \hdr.

\mypar{Host DR} 
\hdr is typically implemented by maintaining a set of round-robin pointers at each source host, one per (destination host, quantized packet size) pair; \hdr uses two quanta bins to triage data and ACK packets. For each pair, the algorithm first determines the lowest common layer in the topology to the source and destination, and performs round-robin among the common switches in this layer using a single pointer. For example, if two hosts are in different pods in \cref{fig:failure-diagram}, they will send their data packets to the four cores in round-robin order. This is done using encapsulation.
When there are failures, \hdr assumes that all hosts have full knowledge of the topology and only performs round-robin among switches that are reachable to both the source and destination hosts. 

\mypar{Bounded queues} We analyze \hdr and show the following results (proved in \cref{sec:DRB}).
Between source and destination hosts, \hdr flows are load-balanced equally and periodically across all allowed paths. As a result, any given path sees a sum of many periodic flows with random phases. Thus, we find that we can model the \hdr queues using the 
$N\,D/D/1$ queueing model~\cite{menth2010packet,roberts1996broadband}. This is significant: unlike most queueing models, \textit{the $N\,D/D/1$ queueing model has bounded average queue sizes even when arrival rates equal service rates}. 
\begin{theorem}
\label{thm:DRB}
The modeled average queue size for \textit{\hdr} is bounded:
\begin{equation} \label{eq:DRB}
    \q \approx \Theta \para{1}.
\end{equation}
\end{theorem}

Given its optimal queue length, we would like to explore a \DR-based load balancing solution. Unfortunately, as we discuss in the next section, \hdr suffers from certain complexities and overheads. This leads us to propose a new variant of DR that we call \textit{\name} which runs in switches and achieves superior performance to \hdr. %

\subsection{Validation via Simulation}
\label{sec:theory-sims}

Since our theoretical analysis makes a number of simplifying assumptions, we present simulation results showing both the CCT and the max queue size   for the various algorithms discussed above. We include both our simplified models and our packet-based leading contenders. For completeness, we also show the performance of \hdr and \name (discussed in the next section).

\begin{figure}
    \centering
    \begin{subfigure}[]{\linewidth} %
        \centering
        \includegraphics[width=\linewidth]{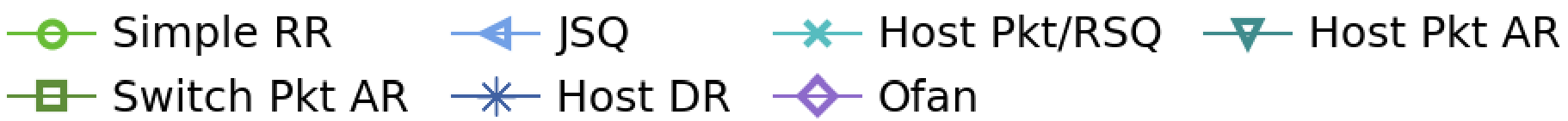}
    \end{subfigure}  
   \par  
    \begin{subfigure}[t]{0.39\linewidth}
        \centering
        \includegraphics[width=\linewidth]{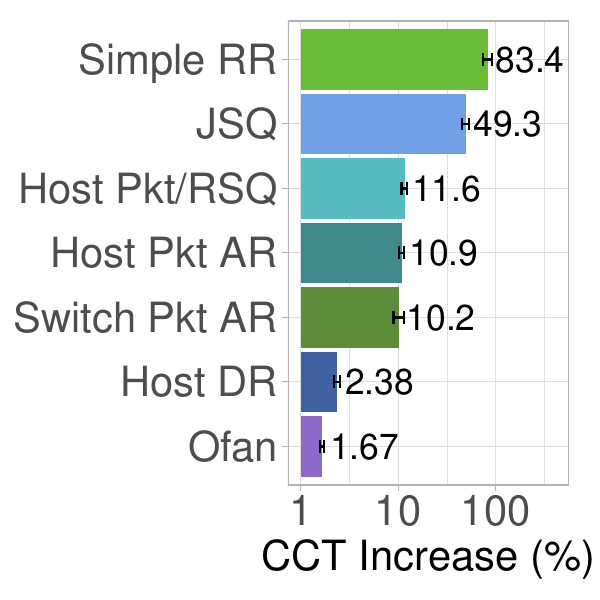}
        \caption{CCT increase}
        \label{fig:spray-cct}
    \end{subfigure}
    \hfill
    \begin{subfigure}[t]{0.57\linewidth}
        \centering
        \includegraphics[width=\linewidth]{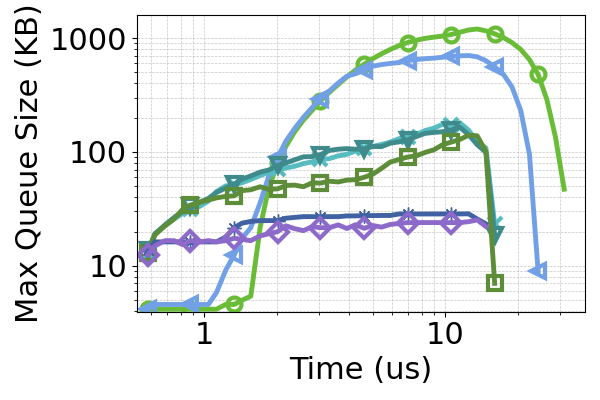}
        \caption{Max queue size}
        \label{fig:max}
    \end{subfigure}
    
    \caption{Comparison of packet-based LB algorithms.}
    \label{fig:growth}
\end{figure}

\Cref{fig:growth} shows that max queue size correlates well with the CCT metric from our earlier sections. We also see the general trends revealed by our theoretical analysis. Specifically, algorithms cluster in three groups: (1) \simplerr and \jsq form a first group with near-linear growth of the max queue size; (2) The second group includes \hpkt, \rsq, \hpktar, and \spktar whose queue growth is closer to the square-root of the message size, and (3) \hdr and \sdr (\name) appear to reach an optimal $O(1)$ queue size.  Finally, we repeated our tests under the failure scenarios from the previous section and confirmed that \name and Host \DR retain their performance advantage over the other LB schemes even under failure (results omitted for brevity). %

Given these results, we next take a closer look at practically realizing \DR-based LB.

\section{\name} \label{sec:algo}

In this section, we first provide intuition for why \DR achieves optimality, then discuss the limitations of \hdr and finally the detailed design of \name. 

\mypar{The reason behind \DR's optimality} Why is destination-based rotation (\DR) optimal, while a scheme based on simple round-robin is not? 
In a nutshell, this is because, by balancing load on a per destination basis, \DR ensures a uniform load on both uplinks \emph{and downlinks}.\footnote{Uplinks refers to the links traversed on a packet's ``northbound'' path from source to a core switch, and downlinks refers to the links traversed on the ``southbound'' path from core to destination.} 
In a scheme such as \simplerr, the switch uses the packet's destination to determine the group of next-hop links, but the destination does not influence \emph{which} of these links is selected. Instead, the switch simply schedules all packets bound for that next-hop group in round-robin order. Or, stated in terms of pointers: the switch only maintains one pointer for each next-hop group. Thus, with \simplerr on a fat-tree, a switch will round-robin all northbound traffic over its uplinks in a destination-agnostic manner. 
This does a good job of balancing traffic across uplinks but can result in imbalances in how \emph{a particular destination's} traffic is spread across the core switches. This in turn can lead to imbalanced traffic on the downlinks because, on a fat-tree, there is only one southbound path from core to destination and hence no further opportunity to load balance traffic.  

By contrast, because \DR explicitly balances load on a per-destination basis, it ensures that \emph{any particular destination}'s traffic is well-balanced across core switches and hence along downlinks, leading to \DR's guarantee of optimality and resilience to cross-flow synchronization effects. The tradeoff, of course, is that \DR requires per-destination pointers and this is the challenge that any \DR-based scheduler must address as we discuss next. 

\mypar{Limitations of \hdr}
\hdr implements \DR at hosts, where maintaining per-destination state is reasonable. 
However, as mentioned before, \hdr suffers from two fundamental limitations. First, hosts must know the complete set of currently reachable switches, which is not information they have available today. While one might contemplate exposing routing/link updates to hosts, doing so would violate traditional network layering abstractions, exposing hosts to the internal complexity of the network fabric and requiring them to manage global topology state.
Second, as we discuss next, running \DR at hosts fundamentally limits the volume of traffic that can be serviced by a single pointer, which in turn limits its performance.

\mypar{Overview of \name} 
\textit{\name}\footnote{\name's name is based on Ezekiel's vision of wheels within wheels, fitting the many pointer rotations.} is a switch-based realization of \DR scheduling. 
To address the challenge of maintaining per-destination pointers, \name exploits the structured nature of a fat-tree topology to \emph{consolidate} destinations based on their location in the tree, such that a switch only maintains a single pointer for groups of destinations that share a common edge or aggregation switch.  %
Specifically, an edge switch $E$ will maintain one pointer for each group of  destinations that share an edge switch $F$ ($F \ne E$). Similarly, an aggregation switch $A$ maintains one pointer for each group of destinations that share an aggregation switch $B$ ($B \ne A$). Hence, an edge switch on the northbound path will round-robin all traffic bound for the same destination edge switch, and an aggregation switch will round-robin traffic bound for the same destination aggregation switch. 

\name's consolidation of destinations retains \DR's property of balancing traffic on both uplinks and downlinks. To see this, consider two flows that originate at source hosts under source edge switch $E$, and are destined to destination hosts under destination edge switch $F$. Then $E$ and $F$ are \textit{mandatory waypoints} in their path and hence $E$ can maintain a single pointer that (effectively) consolidates both flows. Consolidation is even more effective at the aggregation layer due to the properties of three-level fat-trees. In a fat-tree, all packets destined for a pod, that come through a source aggregation switch in their uplink path can only go through a single aggregation switch in their downlink path.\footnote{To see this, notice that any packets leaving the black aggregation switch in Figure~\ref{fig:failure-diagram} can only traverse the red (and not white) aggregation switches on their southbound path.} This aggregation switch is again a mandatory waypoint and hence a source aggregation switch can consolidate its scheduling of all packets destined to this pod. %

Thus, an edge switch maintains O(\# edge switches) pointers, and an aggregation switch maintains O(\# pods) pointers. While greater than the number of pointers maintained with \simplerr, our discussions with switch vendors suggest that this is a very reasonable number of pointers for switches to maintain. For example, in a fat-tree of parameter $\k=64$ with $65K$ GPUs~\cite{metaDC,nvidiaDC,broadcomSwitch,nvidiaSwitch}, each edge switch would maintain $2K$ pointers and each aggregation switch $63$ pointers.

\mypar{Properties} We prove in \cref{sec:ofan} that \name's consolidation achieves the dual goals of uplink and downlink LB. We get:
\begin{theorem}
\label{thm:ofan}
The modeled average queue size for \textit{\name} is bounded:
\begin{equation} \label{eq:ofan}
    \q \approx \Theta \para{1}.
\end{equation}
\end{theorem}

\begin{figure}
    \centering
    \begin{subfigure}[]{0.65\columnwidth}
        \centering
        \includegraphics[width=\linewidth]{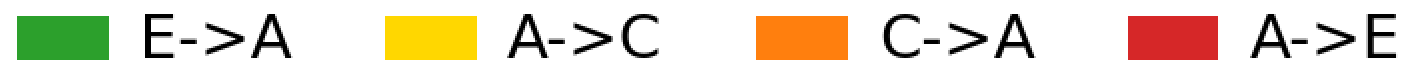}
    \end{subfigure}%
    \par 
    \begin{subfigure}[t]{0.49\columnwidth}
        \centering
        \includegraphics[width=\linewidth]{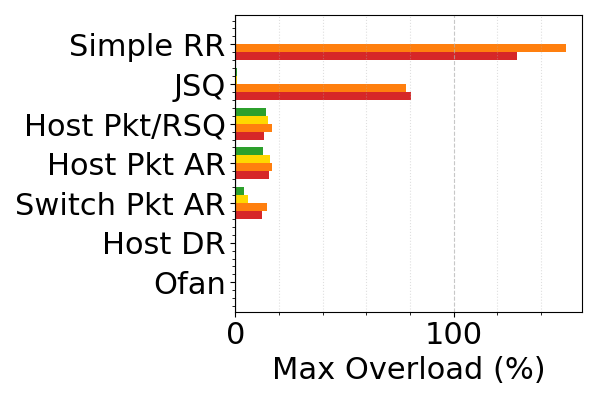}
        \caption{$1\unit{MB}$ message}
        \label{fig:o1MB}
    \end{subfigure} 
    \hfill
    \begin{subfigure}[t]{0.49\columnwidth}
        \centering
        \includegraphics[width=\linewidth]{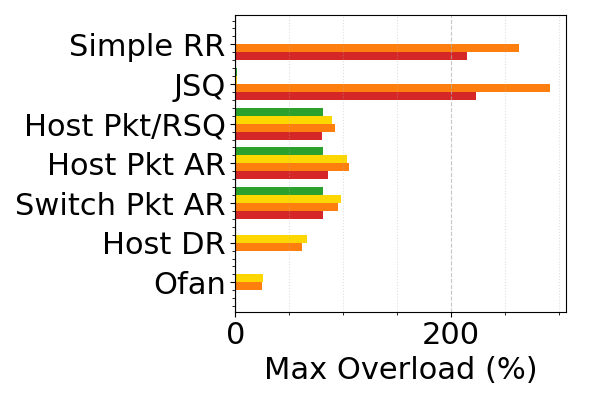}
        \caption{$32\unit{KB}$ message}
        \label{fig:o16KB}
    \end{subfigure} 
    \caption{Worst-case link overload at each stage as a function of packet size for $1\unit{MB}$ and small $32\unit{KB}$ messages. %
    }
    \label{fig:overload}
    \vspace{-0.5em}
\end{figure}

\mypar{Implementation}
Implementing \name only requires modifying the core scheduling algorithm in a switch. No change is required at endhosts and \name can continue to use existing routing protocols such as BGP W-ECMP~\cite{ietf-bess-ebgp-dmz-08,ietf-idr-link-bandwidth-24,rfc7938, nokia-w-ecmp,nvidia-w-ecmp, juniper-w-ecmp,arista-w-ecmp,huawei-w-ecmp} to propagate information about failures \etc 
To implement the \name scheduling algorithm, an edge switch maintains a pointer for each (destination edge switch, quantized packet size) tuple and an aggregation switch maintains the same for each destination pod. In our simulations, we use two quantization groups for packet sizes---one for data packets and another for ACKs. Other datacenters and workloads might require more fine-grained quantization.  At start up, each pointer is initialized to a random uplink port and a random round-robin traversal order (to avoid synchronization across pointers).
At each packet arrival for this pointer, it is incremented (modulo the number of uplink ports). Unlike \hdr, \name does not require any tunneling or encapsulation of traffic.

\mypar{Experimental Validation}
Figure~\ref{fig:overload} provides experimental evidence for why \DR-based LB performs better. 
We plot the maximum load increase beyond ideal at each of the four inter-switch link layers:
(i)~edge to aggregation ($E\to A$),   
(ii)~aggregation to core ($A\to C$),   
(iii)~core to aggregation ($C\to A$), and
(iv)~aggregation to edge ($A\to E$), using a random permutation with inter-pod traffic only.
The results confirm our intuition. \simplerr and \jsq achieve perfect local LB at uplinks, but since they are not destination-based, they suffer at downlinks. The other non-\DR algorithms suffer at both. 
Interestingly, we see that \hdr exhibits perfect LB for large messages, but suffers for small ones. This is because \hdr maintains many per-flow pointers that can cause traffic to be ``spread too thin'', 
especially with small messages. \name's consolidation mitigates this, leading to improved performance. In \cref{sec:ofan}, we show that this performance gap widens as the network size grows, following the widening disparity in the number of pointers maintained by each scheme.

\section{Sensitivity Analysis}
\label{sec:sensitivity}

We now determine the robustness of our results when varying several scaling parameters (\S\ref{sec:scaling}),  approaches to loss recovery (\S\ref{sec:loss}), and congestion control (\S\ref{sec:cca}). 
Finally, we present results measuring CCT in a scenario simulating the training of Llama~\cite{llama3modelcard} models with up to 405B parameters parallelized using FSDP~\cite{rajbhandari2020zero,zhao2023pytorch} and running a full stack of CCA, loss recovery, and load balancing (\S\ref{sec:e2e}). 

We ran tests with the entire set of our leading contenders from \S\ref{sec:space} as well as \hdr and \name. Across our tests, we find that the \emph{relative} ranking and broad assessment of  different schemes that we have made so far remains unchanged. For legibility, in this section, we show results for \spktar, \hpktar, and \name since these are our best performers representing both practice and theory.

\subsection{Scaling}
\label{sec:scaling}

\begin{figure}
    \centering
    \begin{subfigure}[]{1\columnwidth}
    \centering
        \includegraphics[width=0.6\columnwidth]{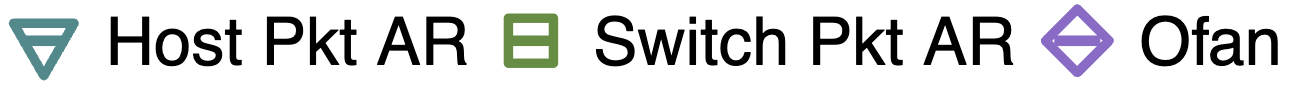}
    \end{subfigure}
    \begin{subfigure}[b]{0.48\columnwidth}
        \includegraphics[width=\columnwidth]{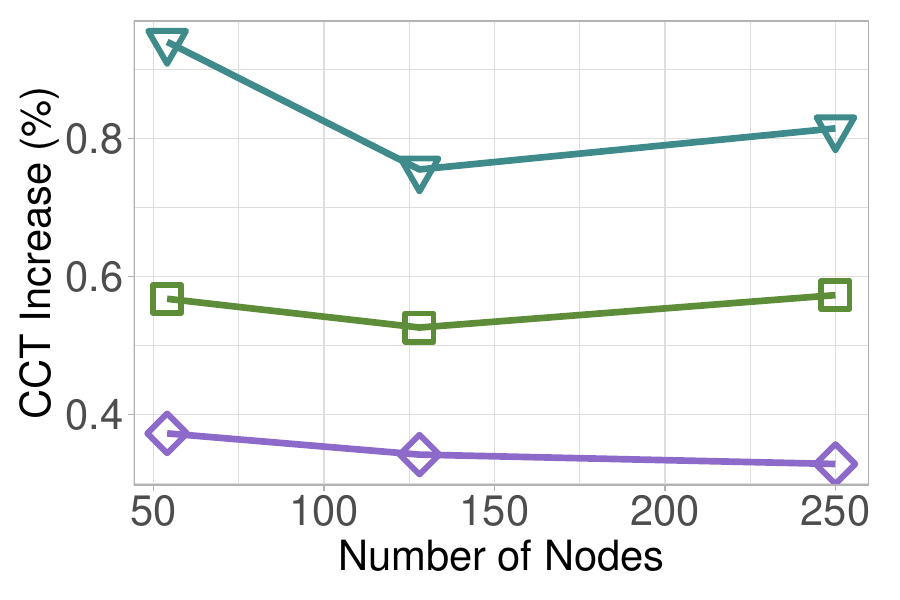}
        \caption{All-to-All}
    \end{subfigure}%
    \hspace{0.02\columnwidth}
    \begin{subfigure}[b]{0.48\columnwidth}
        \includegraphics[width=\columnwidth]{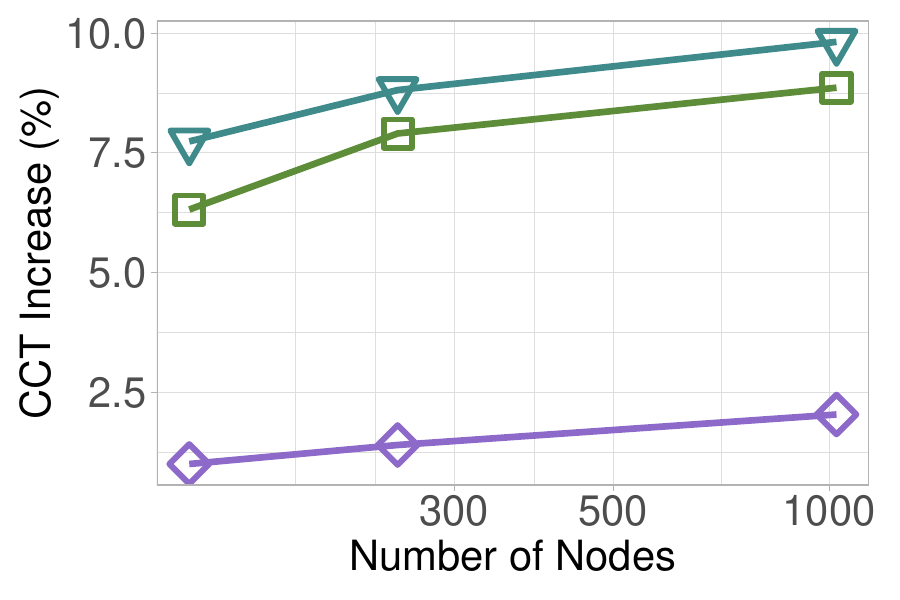}
        \caption{Permutation}
    \end{subfigure}
    \caption{ CCT increase as the network size increases.}
    \label{fig:networksize}
    \vspace{-0.5em}
\end{figure}

\mypar{Network Size}
\Cref{fig:networksize} shows the impact of scaling the number of nodes in the network (and thus participating in the collective). 
We see that both the relative ordering of our schemes and the magnitude of their performance differences holds as the network size grows.

\begin{figure}
    \centering
    \begin{subfigure}[b]{0.48\columnwidth}
        \includegraphics[width=\columnwidth]{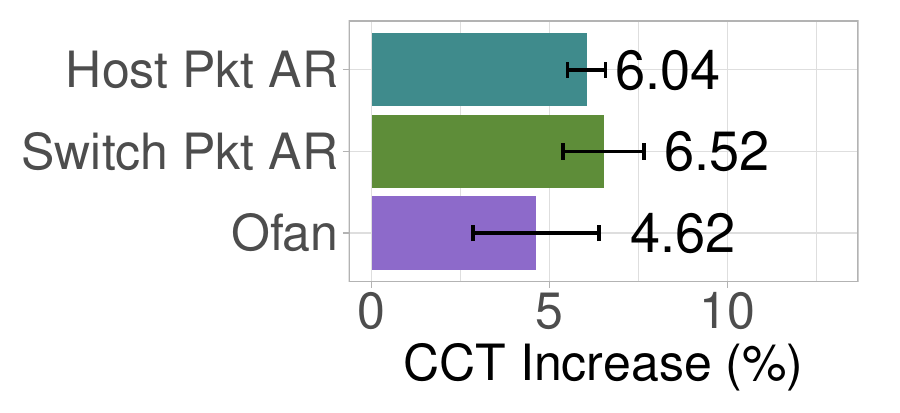}
        \caption{All-to-All}
    \end{subfigure}%
    \begin{subfigure}[b]{0.48\columnwidth}
        \includegraphics[width=\columnwidth]{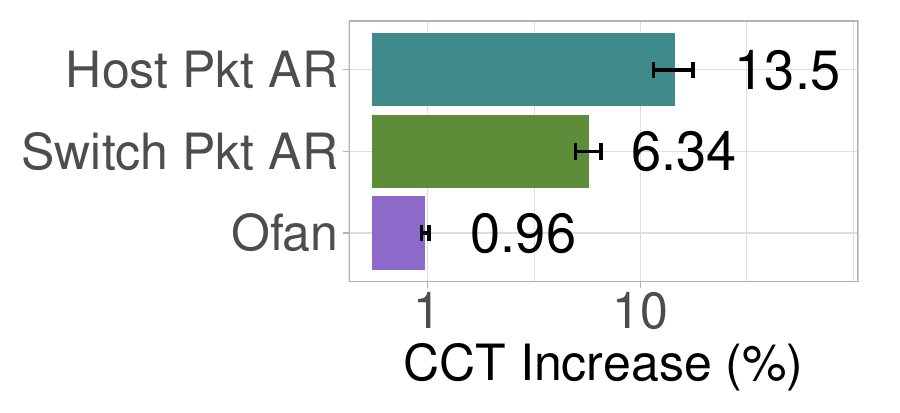}
        \caption{Permutation}
    \end{subfigure}
    \caption{ CCT increase with short (20 pkt) buffers. }
    \label{fig:buffersize}
\end{figure}

\mypar{Buffer Size}
In \Cref{fig:buffersize}, we show performance with short buffers, set at the size of 20 data packets (1/10th of the default in \S\ref{sec:lb}). We see that, for permutation matrices, the performance of \spktar starts to dominate Host Pkt AR, presumably because of its ability to more directly control queue length. 
As before, \name consistently performs better than the other approaches.

\begin{figure}
    \centering
    \begin{subfigure}[]{1\columnwidth}
    \centering
        \includegraphics[width=0.6\columnwidth]{figures26/sweep-legend.png}
    \end{subfigure}
    \begin{subfigure}[b]{0.48\columnwidth}
        \includegraphics[width=\columnwidth]{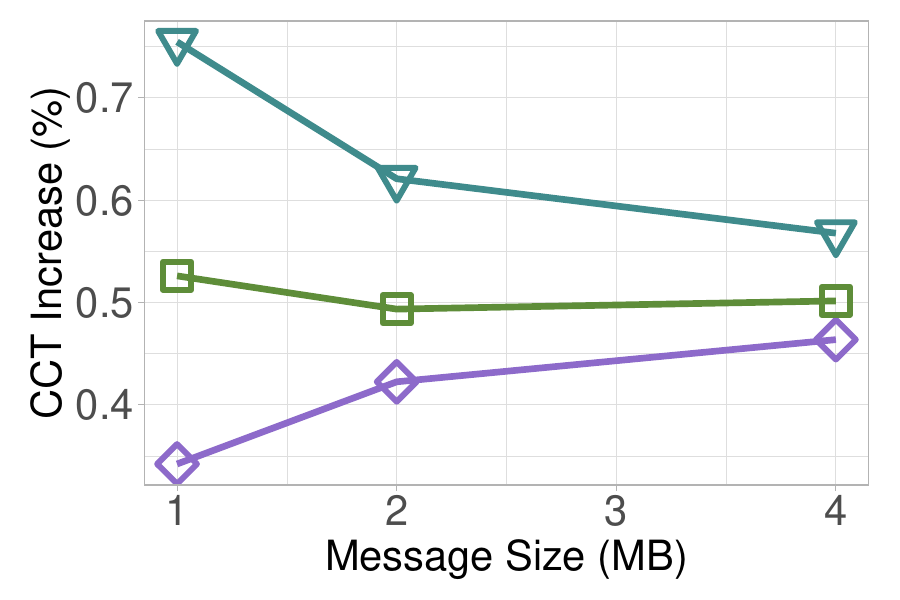}
        \caption{All-to-All}
    \end{subfigure}%
    \hspace{0.02\columnwidth}
    \begin{subfigure}[b]{0.48\columnwidth}
        \includegraphics[width=\columnwidth]{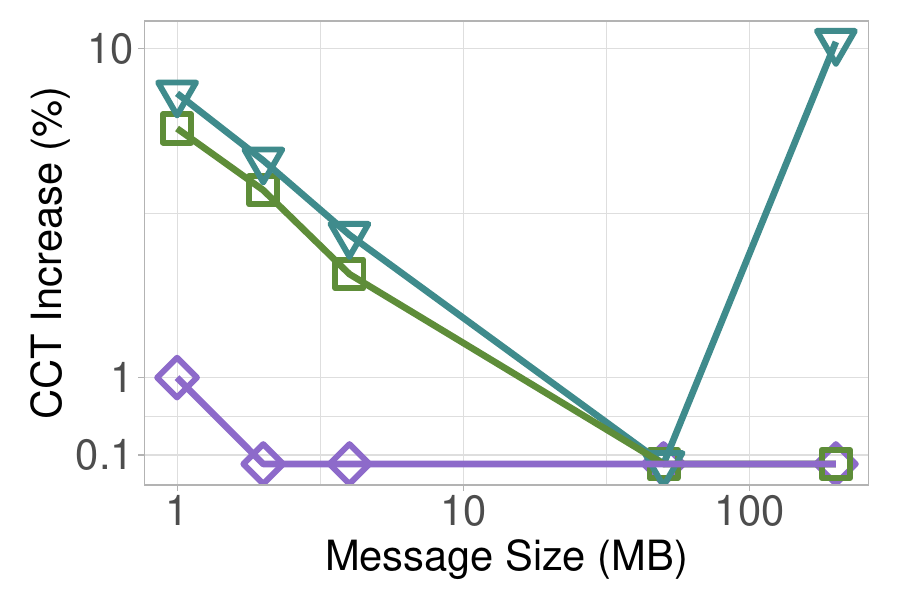}
        \caption{Permutation}
    \end{subfigure}
    \caption{CCT increase as the message size  increases.}
    \label{fig:messagesize}
\end{figure}

\mypar{Message Size}
\Cref{fig:messagesize} plots performance under increasing message size. We see that the relative ordering of our approaches  does not significantly change, though the gap between them decreases. 
This is because, as the message size increases, the minimum CCT increases and hence the relative differences get smaller. %
Interestingly, we see that \hpktar gets notably worse at very large message size. We conjecture that this is due to the ``freezing mode'' in REPS triggering as queues become long enough to cause a timeout~\cite{bonatoreps}. %

\begin{figure}
    \centering
    \begin{subfigure}[]{\columnwidth}
        \includegraphics[width=\columnwidth]{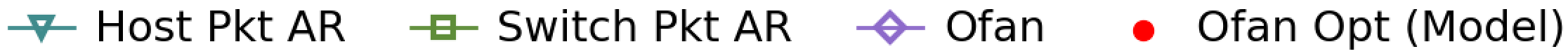}
    \end{subfigure}    
    \begin{subfigure}[t]{0.48\columnwidth}
        \centering
        \includegraphics[width=\linewidth]{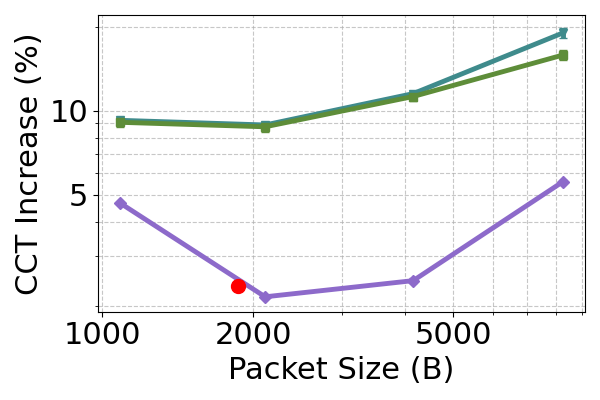}
        \caption{$0.5\unit{MB}$ message}
        \label{fig:pktsize1}
    \end{subfigure}
    \hfill
    \begin{subfigure}[t]{0.48\columnwidth}
        \centering
        \includegraphics[width=\linewidth]{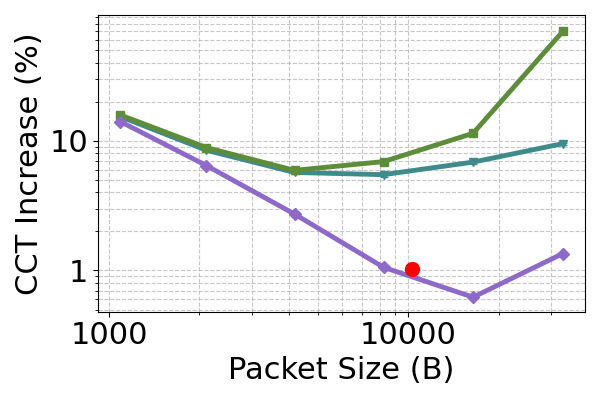}
        \caption{$16\unit{MB}$ message}
        \label{fig:pktsize2}
    \end{subfigure}
    \caption{CCT increase as packet size increases, for two message sizes in bytes.
    }
    \label{fig:pktsize}
\end{figure}

\mypar{Packet Size} 
\Cref{fig:pktsize} repeats our default test, now with varying packet size and two different message sizes. The impact of packet size as a parameter in AI training is one that, to our knowledge, has not been extensively explored in the networking literature. Yet, as we see, the choice of packet size can have a significant impact, varying CCT by up to 10\%. The graph also shows a clear ``sweet spot'' in packet size at which performance is maximized. 
Intuitively, this is because sending $D$ bytes with a packet size $P$ and header $H$, involves a trade-off: large packets increase queue size and hence CCT, while small packets incur higher header overheads. We find that we can model the optimal packet size for some of our LB approaches. 
For \DR schemes such as \name, if we model the  $O(1)$ network queueing as a constant of $\alpha$ packets 
(independent of the number of sent packets), 
then we obtain the following (\cref{sec:pkt}):

\begin{theorem}
\label{thm:pkt}
To minimize the CCT, the  optimal packet size should carry a payload that scales with $\sqrt{D}$:
\begin{equation} \label{eq:payload}
P-H = \sqrt{\frac{H}{\alpha} \cdot D}
\end{equation}
\end{theorem}

Interestingly, the constant $H/\alpha$ reflects the battle between the two types of CCT overhead: \textit{headers} and \textit{queueing}.
We illustrate with a red dot in~\cref{fig:pktsize} the optimal size and modeled CCT increase \vs our best constructed lower bound (\cref{sec:lb}). We can see how the model roughly predicts the packet-size sweet spot observed in simulation and its CCT increase. 
\cref{fig:pktsize2} shows how the shorter queue sizes of \name also encourage larger packets: We can derive that when using spraying algorithms with square-root queue-size growth, optimal packet size only grows as $\Theta\para{D^{1/3}}$ rather than $\Theta\para{\sqrt{D}}$. 
Also note that we neglect second-order factors, \eg the CRC may need to be a bit longer to protect longer packets.
This result suggests many avenues for future work on optimizing performance through dynamically tuning packet size, and exploring standardization support for larger Ethernet MTUs.

\subsection{Loss Recovery}
\label{sec:loss}
\begin{figure}
    \centering
    \begin{subfigure}[b]{0.48\columnwidth}
        \includegraphics[width=\columnwidth]{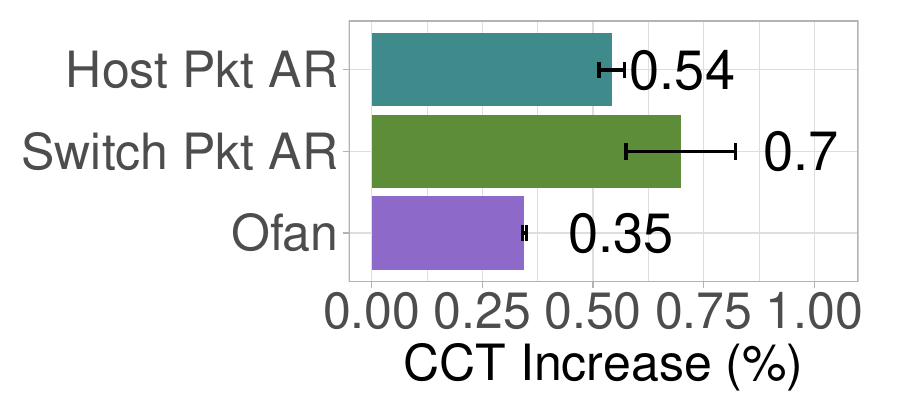}
        \caption{All-to-All}
        \label{subfig:ata-sack}
    \end{subfigure}%
    \begin{subfigure}[b]{0.48\columnwidth}
        \includegraphics[width=\columnwidth]{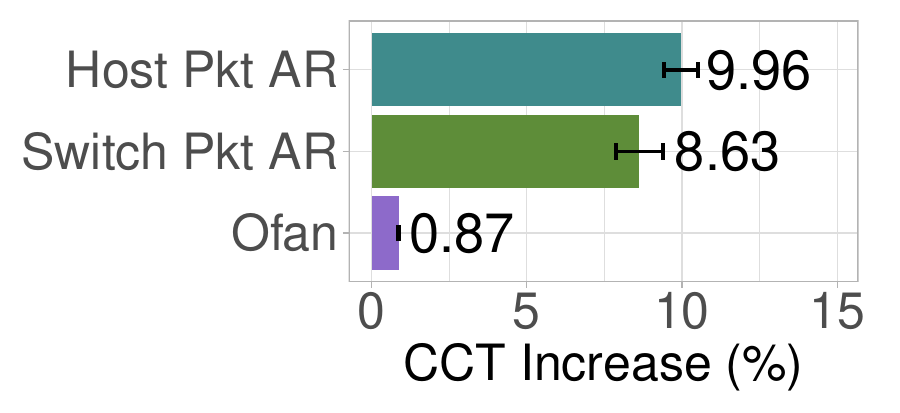}
        \caption{Permutation}
        \label{subfig:perm-sack}
    \end{subfigure}
    \caption{CCT increase for each LB scheme with SACK-based loss recovery protocol.}
    \label{fig:sack}
\end{figure}

\begin{figure}
\centering
\begin{subfigure}{0.49\linewidth}
	\includegraphics[width=\textwidth]{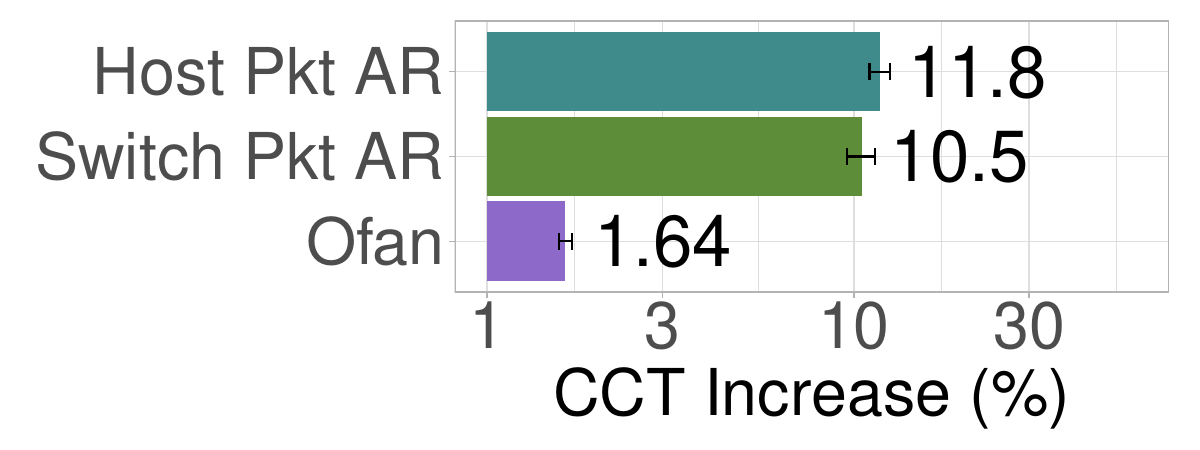}
	\caption{$1\unit{MB}$: CCT }%
	\label{fig:cca0}
\end{subfigure}
\hfill
\begin{subfigure}{0.49\linewidth}
	\includegraphics[width=\textwidth]{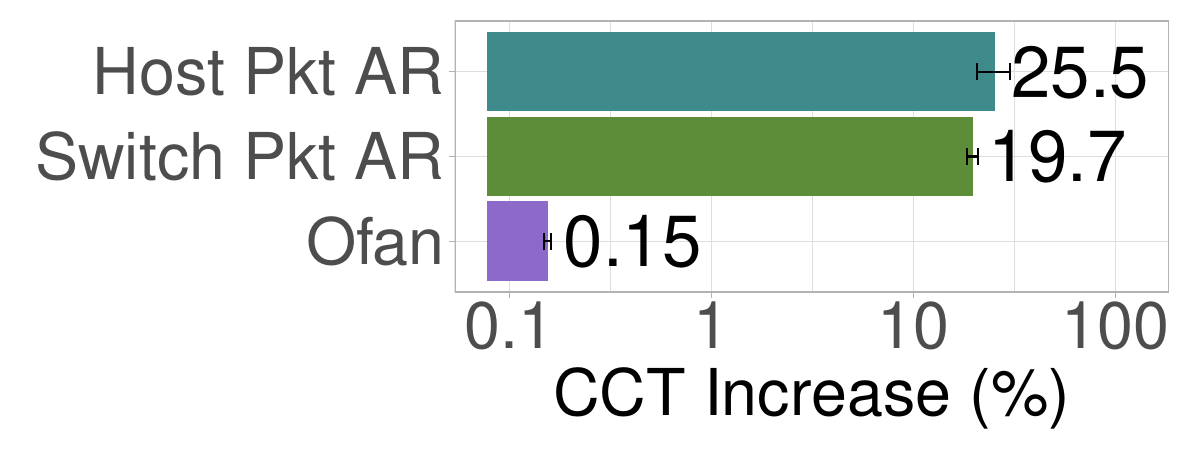}
	\caption{$16\unit{MB}$: CCT
    }%
	\label{fig:cca2}
\end{subfigure}
\caption{Packet-based LB algorithms with MSwift CCA.
}
\label{fig:cca}
\end{figure}

We next evaluate the impact of a realistic loss recovery mechanism on CCT, using a retransmission threshold $x$ to distinguish expected reordering from actual loss. Analytically, the maximum reordering degree -- the packets arriving while one is delayed -- is $5qr/b$ for a 3-level fat-tree with queue size $q$, bandwidth $b$, and flow rate $r$ (\ie the maximum possible queueing delay divided by the pacing time between packets). In our setup, this worst-case threshold ranges from 8 packets for ATA to 1,000 for the permutation matrix. 

Empirical sweeps identified optimal thresholds of $x=6$ for ATA and $x=32$ for the permutation matrix (avoiding spurious retransmissions due to reordering, but not relying entirely on timeouts to detect loss). As shown in \cref{fig:sack}, incorporating realistic loss recovery does not alter our core findings: existing spraying approaches perform similarly, while \name consistently provides the best performance across both traffic matrices.

\begin{figure*}
    \centering
    \begin{subfigure}[h!]{0.35\linewidth}
        \includegraphics[width=\linewidth]{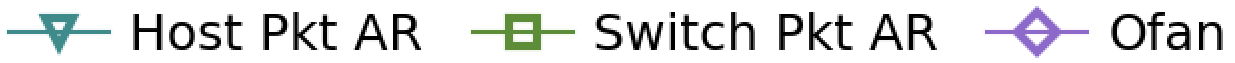}
    \end{subfigure}  
    \par    
    \begin{subfigure}[t]{0.17\linewidth}
        \centering
        \includegraphics[width=\linewidth]{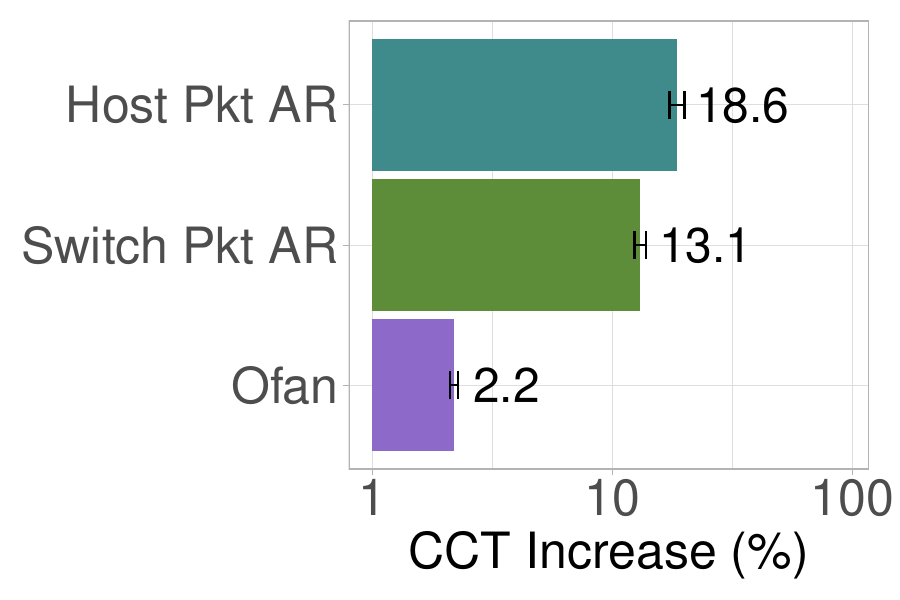}
        \caption{7B: CCT increase}
        \label{fig:F1}
    \end{subfigure}
    \hfill
    \begin{subfigure}[t]{0.17\linewidth}
        \centering
        \includegraphics[width=\linewidth]{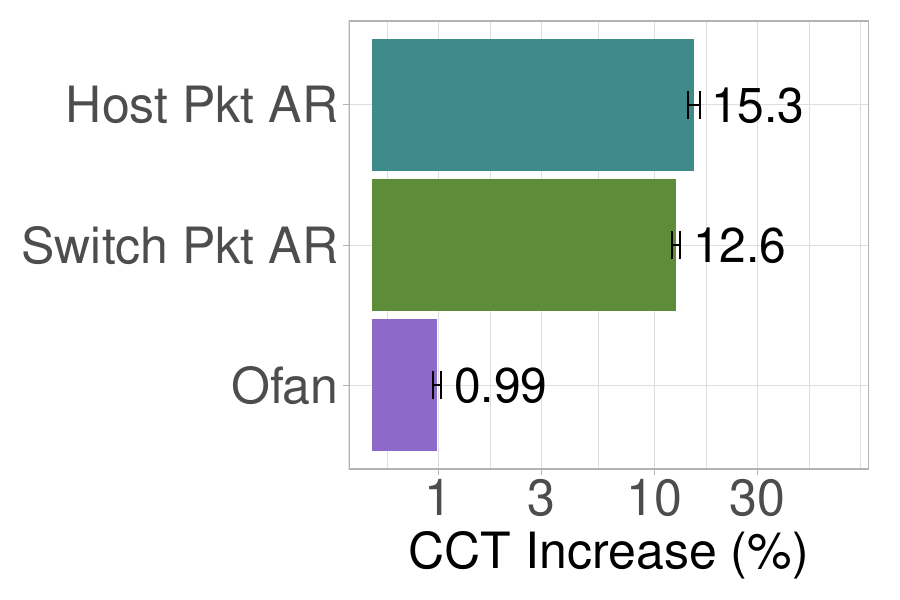}
        \caption{70B: CCT increase}
        \label{fig:F2}
    \end{subfigure}
    \hfill
    \begin{subfigure}[t]{0.17\linewidth}
        \centering
        \includegraphics[width=\linewidth]{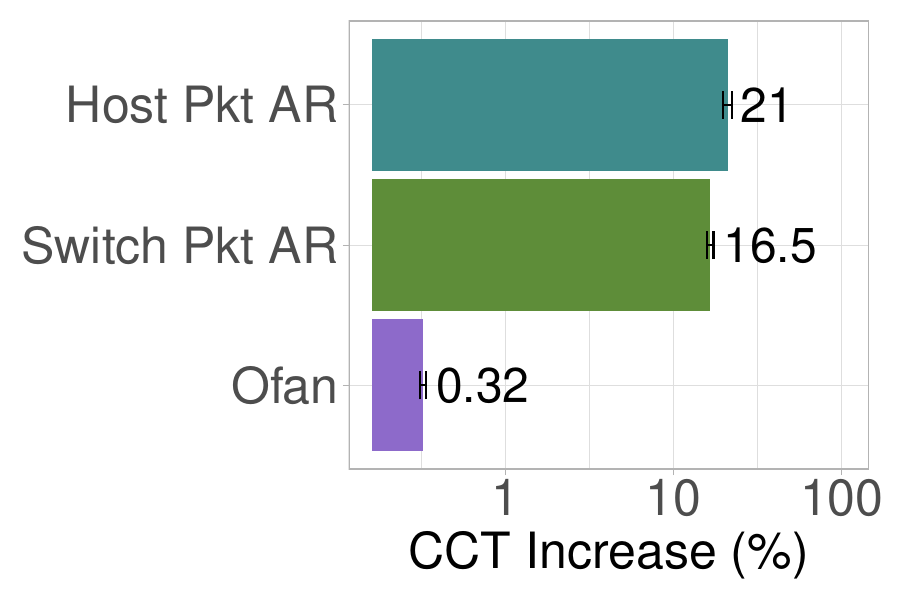}
        \caption{405B: CCT increase}
        \label{fig:F3}
    \end{subfigure}
    \hfill    
    \begin{subfigure}[t]{0.22\linewidth}
        \centering
        \includegraphics[width=\linewidth]{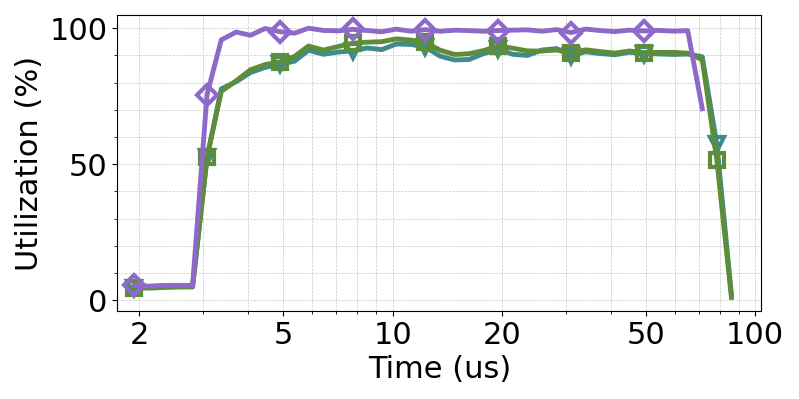}
        \caption{405B: Utilization}
        \label{fig:F4}
    \end{subfigure}    
    \hfill
    \begin{subfigure}[t]{0.22\linewidth}
        \centering
        \includegraphics[width=\linewidth]{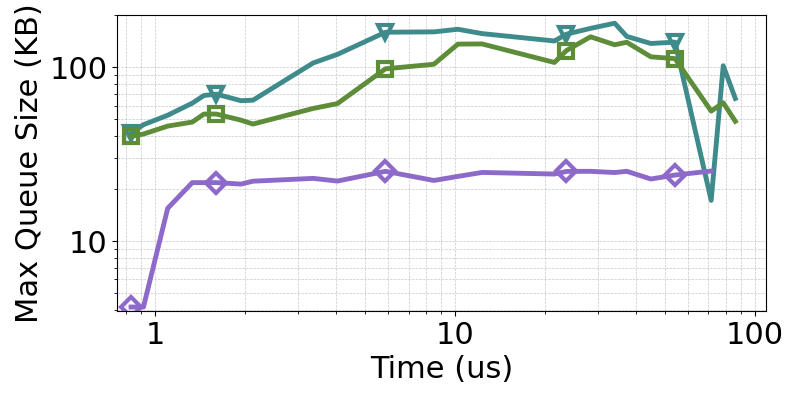}
        \caption{405B: Max queue size}
        \label{fig:F5}
    \end{subfigure}
    \caption{FSDP with Llama 7B, 70B and 405B parameters.}
    \label{fig:FSDP}
\end{figure*}

\subsection{Congestion Control }
\label{sec:cca}

We now compare per-packet spraying algorithms with CCA. 
As mentioned previously, we would like to compare them using Google's Swift~\cite{swift}, but Swift is designed for single paths and experiences throughput collapse when handling spraying~\cite{mswift}. Therefore, we use MSwift~\cite{mswift}, a recent and publicly available CCA that extends Swift to operate under spraying. 
MSwift's key changes include: (i)~an LTCP-based mechanism~\cite{ltcp} to handle out-of-order packets that can trigger duplicate ACKs, and (ii)~a new mechanism (based on observing recent ACKs) that avoids Swift’s throughput collapse scenario. We use MSwift's default parameter values adjusted for our network setting~(\cref{sec:swift}).

\Cref{fig:cca} illustrates the impact of congestion control. It evaluates a permutation traffic using the same settings as without CCA, both with a short $1\unit{MB}$ message to compare with \cref{fig:growth}, and with a longer $16\unit{MB}$ one to see the influence of the CCA over several RTTs. For a short message, starting with congestion windows reflecting full line rates, \Cref{fig:cca0} predictably shows largely similar CCT results. 
\Cref{fig:cca2} shows that with longer messages, the CCA needs to rein down the flow rates, and therefore the CCT inflation can worsen by some 10\% for \hpktar and \spktar.
By contrast, \name does not need to be slowed down, reducing its CCT by an order of magnitude. Intuitively, the CCT inflation is the ratio of a variable queueing delay by a CCT lower bound. The numerator stays near constant because of the $\Theta(1)$ queues, while the denominator grows by an order of magnitude with the message size (\ie a bit less than $16\times$ because of the fixed RTT tax).

\subsection{Training Scenario}
\label{sec:e2e}

Finally, we evaluate the performance of each LB scheme on a realistic AI training scenario. 
Specifically, we simulate training Llama-3~\cite{llama3modelcard} (7B, 70B, and 405B) parallelized using Fully Sharded Data Parallelization (FSDP)~\cite{rajbhandari2020zero,zhao2023pytorch} on a 1,024-GPU cluster.
The cluster is organized as 128 servers of 8 GPUs, and we assume random server placement. 
We evaluate the main network bottleneck in FSDP training by focusing on the backward pass, as it combines a ReduceScatter of gradients and an AllGather of weights.
FSDP traffic is implemented as a hierarchical ring, where intra-server communications are performed in the high-bandwidth domain, while our fat-tree network implements inter-server communications using 8 parallel rings: each logical GPU $i$ is connected to logical GPU $i+8 \pmod{1,024}$.
We assume FP8 precision and $4\unit{KB}$ packet payload, translating to 104 packets per flow in Llama 7B (32 layers), 418 in Llama 70B (80 layers) and 1570 in Llama 405B (126 layers). 
We enable the MSwift CCA with loss recovery, and  
measure the CCT of each collective phase.

As shown in \Cref{fig:FSDP}, while \hpktar and \spktar overheads remain constant, the relative CCT increase for \name drops as model size scales. At 405B parameters, \name reduces the CCT increase by $52\times$ over baselines. This improvement stems from superior utilization (\cref{fig:F4}): Since the CCA must throttle \hpktar and \spktar to maintain stable queue sizes (\cref{fig:F5}), \name achieves significantly higher effective throughput.

\section{Related Work}
\label{sec:related}

There is a vast literature on load balancing in datacenters, and we cannot hope to summarize it in any depth. The citations in Table~\ref{tab:designspace} provide a subset that covers the space along the dimensions we described in  \S\ref{sec:space}. Many of the individual papers evaluate specific designs (that make choices for each of the components), but we are not aware of work that tries to systematically evaluate each component in designing a solution. This is especially true for the unique setting of large AI training workloads. 
In addition, many of these proposed systems \cite{letitflowlet, kandulaflowlet, plb, meta} do not seriously consider packet spraying given its potential adverse effects on the transport due to reordering. Some systems do include spraying in their design \cite{ndp, eqds, dcpim,pfabric}, but often introduce specialized mechanisms to make spraying work (\eg proposing an entirely new transport). Two exceptions are \cite{rps} and \cite{bonatoreps}, where packet spraying is evaluated against ECMP and some other load balancing granularities, but one transport is assumed and the distance from optimality is unclear. 
We note that there are a few works that use topology-aware (sub)flow-based load balancing~\cite{ethereal, alibaba} implemented in the collective communication library (CCL), in which subflows are placed on specific paths. 
This contrasts with our evaluated subflow setup as we do not model topology-awareness managed at the host. 
We categorize these into an approach similar to traffic engineering (rather than transparent network or transport-level LB), a space that we leave to exploration in future work. %

\section{Conclusion}

As AI training demands push network fabrics toward packet-level spraying, load-balancing architecture becomes critical. 
We systematically evaluated host and switch-based schemes, finding they perform comparably except when routing convergence is slow. 
However, no leading algorithm achieves optimal $O(1)$ queue scaling at full utilization. 
To close this gap, we introduced \name, a switch-based implementation of destination-based rotation that reaches theoretical queue length optimality.
Our evaluations show \name significantly improves completion times and ensures predictable performance for next-generation AI workloads.

\section*{Acknowledgments}
We thank Nandita Dukkipati, Ziming Mao and David Tennenhouse for their helpful comments. This work was partly supported by the Louis and Miriam Benjamin Chair in Computer-Communication Networks.

\bibliographystyle{ACM-Reference-Format}
\bibliography{references}

@string{tnet = "IEEE/ACM Transactions on Networking"}

@STRING{sigcomm         = "{ACM SIGCOMM}"}

@string{nsdi = "USENIX NSDI"}

@STRING{conext         = "{ACM CoNEXT}"}

@STRING{infocom         = "{IEEE Infocom}"}

@string{eurosys = "ACM EuroSys"}

@article{padded-frames,
    author = {Jaramillo, J.J. and Milan, F. and Srikant, R.},
    title = {Padded frames: A novel algorithm for stable scheduling in load-balanced switches},
    journal = tnet,
    volume = {16},
     number = {5},
    pages = {1212--1225},
    year = {2008}
}

@book{thesis,
  title={The load-balanced router},
  author={Keslassy, Isaac},
  year={2004},
  publisher={Stanford University}
}

@article{LBR,
  author    = {Isaac Keslassy and
               Shang-Tse Chuang and
               Kyoungsik Yu and
               David Miller and
               Mark Horowitz and
               Olav Solgaard and
               Nick McKeown},
  title     = {Scaling internet routers using optics},
  journal = sigcomm,
  year      = {2003},
  volume       = {33},
   number       = {4},
  pages     = {189--200},
  ee        = {http://doi.acm.org/10.1145/863955.863978},
  bibsource = {DBLP, http://dblp.uni-trier.de}
}

@article{menth2010packet,
  title={Packet waiting time for multiplexed periodic on/off streams in the presence of overbooking},
  author={Menth, Michael and Muehleck, Stefan},
  journal={International Journal of Communication Networks and Distributed Systems},
  volume={4},
  number={2},
  pages={207--229},
  year={2010},
  publisher={Inderscience Publishers}
}

@book{roberts1996broadband,
  title={Broadband network traffic: Performance evaluation and design of broadband multiservice networks},
  author={Roberts, James and Mocci, Ugo and Virtamo, Jorma},
  year={1996},
  publisher={Springer}
}

@article{llama3modelcard,
  title={The Llama 3 Herd of Models},
  author={Llama Team, AI @ Meta},
  journal={arXiv preprint arXiv:2407.21783},
  year={2024}
}

@inproceedings{rajbhandari2020zero,
  title={{ZeRO}: Memory optimizations toward training trillion parameter models},
  author={Rajbhandari, Samyam and Rasley, Jeff and Ruwase, Olatunji and He, Yuxiong},
  booktitle={SC20: International Conference for High Performance Computing, Networking, Storage and Analysis},
  pages={1--16},
  year={2020},
  organization={IEEE}
}

@article{zhao2023pytorch,
  title={{PyTorch FSDP}: Experiences on scaling fully sharded data parallel},
  author={Zhao, Yanli and Gu, Andrew and Varma, Rohan and Luo, Liang and Huang, Chien-Chin and Xu, Min and Wright, Less and Shojanazeri, Hamid and Ott, Myle and Shleifer, Sam and others},
  journal={Proceedings of the VLDB Endowment},
  volume={16},
  number={12},
  pages={3848--3860},
  year={2023},
  publisher={VLDB Endowment}
}

@inproceedings{gill2011understanding,
  title={Understanding network failures in data centers: measurement, analysis, and implications},
  author={Gill, Phillipa and Jain, Navendu and Nagappan, Nachiappan},
  booktitle={Proceedings of the ACM SIGCOMM 2011 Conference},
  pages={350--361},
  year={2011}
}

@article{addie2002approximation,
  title={An approximation for performance evaluation of stationary single server queues},
  author={Addie, Ronald G and Zukerman, Moshe},
  journal={IEEE Transactions on Communications},
  volume={42},
  number={12},
  pages={3150--3160},
  year={2002},
  publisher={IEEE}
}

@article{ltcp,
  title={Improving datacenter throughput and robustness with {Lazy TCP} over packet spraying},
  author={Zhang, Jie and Zhang, Dafang and Huang, Kun},
  journal={Computer Communications},
  volume={62},
  pages={23--33},
  year={2015},
  publisher={Elsevier}
}

@article{mswift,
  title={Making congestion control robust to per-packet load balancing in datacenters},
  author={Gerstein, Barak and Silberstein, Mark and Keslassy, Isaac},
  journal={arXiv preprint arXiv:2509.07907},
  year={2025}
}

@techreport{ietf-bess-ebgp-dmz-08,
    number =    {draft-ietf-bess-ebgp-dmz-08},
    type =      {Internet-Draft},
    institution =   {Internet Engineering Task Force},
    publisher = {Internet Engineering Task Force},
    note =      {Work in Progress},
    url =       {https://datatracker.ietf.org/doc/draft-ietf-bess-ebgp-dmz/08/},
    author =    {Stephane Litkowski and Satya R Mohanty and Arie Vayner and Akshay Gattani and Ajay Kini and Jeff Tantsura and Reshma Das},
    title =     {{BGP link bandwidth extended community use cases}},
    pagetotal = 17,
    year =      2025,
    month =     oct,
    day =       17,
}

@article{katevenis1991weighted,
  title={Weighted round-robin cell multiplexing in a general-purpose ATM switch chip},
  author={Katevenis, Manolis and Sidiropoulos, Stefanos and Courcoubetis, Costas},
  journal={IEEE Journal on selected Areas in Communications},
  volume={9},
  number={8},
  pages={1265--1279},
  year={1991},
  publisher={IEEE}
}

@article{tabatabaee2021interleaved,
  title={Interleaved weighted round-robin: A network calculus analysis},
  author={Tabatabaee, Seyed Mohammadhossein and Le Boudec, Jean-Yves and Boyer, Marc},
  journal={IEICE Transactions on Communications},
  volume={104},
  number={12},
  pages={1479--1493},
  year={2021},
  publisher={The Institute of Electronics, Information and Communication Engineers}
}

@misc{nvidiaSwitch,
  title = {{NVIDIA® 64-Port NDR 400G InfiniBand Data Center Switch}},
  author={{FS}},
  year  = {2026},
  howpublished={\url{https://www.fs.com/products/194710.html}}
  }

@misc{broadcomSwitch,
  title = {{Tomahawk4 / BCM56990 Series}},
  author={{Broadcom}},
  year  = {2026},
  howpublished={\url{https://www.broadcom.com/products/ethernet-connectivity/switching/strataxgs/bcm56990-series}}
  }

@misc{nvidiaDC,
  title = {{NVIDIA DGX SuperPOD: Next Generation Scalable Infrastructure for AI Leadership Reference Architecture Featuring NVDIA DGX H100
}},
  author={{NVIDIA}},
  year  = {2026},
  howpublished={\url{https://docs.nvidia.com/dgx-superpod/reference-architecture-scalable-infrastructure-h100/latest/components.html}}
  }

@misc{metaDC,
  title = {{Building Meta’s GenAI Infrastructure}},
  author={{Lee, Kevin and Gangidi, Adi and   Oldham, Mathew}},
  year  = {2024},
  howpublished={\url{https://engineering.fb.com/2024/03/12/data-center-engineering/building-metas-genai-infrastructure/}}
  }

@misc{nokia-w-ecmp,
  title = {{BGP}},
  author={{NOKIA}},
  year  = {2026},
  howpublished={\url{https://documentation.nokia.com/srlinux/25-7/books/routing-protocols/bgp.html}}
  }

@misc{juniper-w-ecmp,
  title = {{BGP Link-Bandwidth Community}},
  author={{HPE Juniper}},
  year  = {2026},
  howpublished={\url{https://www.juniper.net/documentation/us/en/software/junos/ai-ml-evo/bgp/topics/topic-map/bgp-link-bandwidth.html}}
  }

@misc{arista-w-ecmp,
  title = {{UCMP}},
  author={{Arista}},
  year  = {2026},
  howpublished={\url{https://www.arista.com/en/support/toi/tag/ucmp}}
  }

@misc{huawei-w-ecmp,
  title = {{CloudEngine 9800, 8800, and 6800 V300R024C10, C11 Configuration Guide - IP Routing}},
  author={{Huawei}},
  year  = {2026},
  howpublished={\url{https://info.support.huawei.com/hedex/api/pages/EDOC1100277644/AEM10221/04/resources/vrp/dc_vrp_bgp_cfg_4094.html}}
  }

@misc{nvidia-w-ecmp,
  title = {{BGP Weighted Equal Cost Multipath}},
  author={{NVIDIA}},
  year  = {2026},
  howpublished={\url{https://docs.nvidia.com/networking-ethernet-software/cumulus-linux-515/Layer-3/Routing/BGP-Weighted-Equal-Cost-Multipath/}}
  }

@techreport{ietf-idr-link-bandwidth-24,
    number =    {draft-ietf-idr-link-bandwidth-24},
    type =      {Internet-Draft},
    institution =   {Internet Engineering Task Force},
    publisher = {Internet Engineering Task Force},
    note =      {Work in Progress},
    url =       {https://datatracker.ietf.org/doc/draft-ietf-idr-link-bandwidth/24/},
    author =    {Prodosh Mohapatra and Reshma Das and SATYA R MOHANTY and Serge Krier and Rafal Jan Szarecki and Akshay Gattani},
    title =     {{BGP Link Bandwidth Extended Community}},
    pagetotal = 12,
    year =      2026,
    month =     jan,
    day =       7,
}

@misc{rfc7938,
    series =    {Request for Comments},
    number =    7938,
    howpublished =  {RFC 7938},
    publisher = {RFC Editor},
    doi =       {10.17487/RFC7938},
    url =       {https://www.rfc-editor.org/info/rfc7938},
    author =    {Petr Lapukhov and Ariff Premji and Jon Mitchell},
    title =     {{Use of BGP for Routing in Large-Scale Data Centers}},
    pagetotal = 35,
    year =      2016,
    month =     aug,
}

@misc{nvidia_AR_whitepaper,
  title = {{NVIDIA InfiniBand Adaptive Routing Technology Accelerating HPC and AI Applications}},
  author={{NVIDIA}},
  year  = {2023},
  howpublished={\url{https://www.amax.com/content/files/2023/12/NVIDIA_InfiniBand_Adaptive_Routing_Technology_Insights_Whitepaper.pdf}}
  }

@misc{nvidia_AR_whitepaper2,
  title = {{NVIDIA Spectrum-X Network
Platform Architecture}},
  author={{NVIDIA}},
  year  = {2024},
  howpublished={\url{https://resources.nvidia.com/en-us-networking-ai/nvidia-spectrum-x}}
  }

@inproceedings{gomez2007deterministic,
  title={Deterministic versus adaptive routing in fat-trees},
  author={Gomez, Crisp{\'\i}n and Gilabert, Francisco and Gomez, Mar{\'\i}a Engracia and L{\'o}pez, Pedro and Duato, Jos{\'e}},
  booktitle={2007 IEEE International Parallel and Distributed Processing Symposium},
  pages={1--8},
  year={2007},
  organization={IEEE}
}

@book{duato650interconnection,
  title={Interconnection Networks: An Engineering Approach. 2002},
  author={Duato, Jose and Yalamanchili, Sudhakar and Lionel, Ni},
  publisher={Morgan Kaufmann Publishers Inc}
}

@inproceedings{drb,
  title={Per-packet load-balanced, low-latency routing for {C}los-based data center networks},
  author={Cao, Jiaxin and Xia, Rui and Yang, Pengkun and Guo, Chuanxiong and Lu, Guohan and Yuan, Lihua and Zheng, Yixin and Wu, Haitao and Xiong, Yongqiang and Maltz, Dave},
  booktitle=conext,
  pages={49--60},
  year={2013}
}

@inproceedings{ndp,
  title={Re-architecting datacenter networks and stacks for low latency and high performance},
  author={Handley, Mark and Raiciu, Costin and Agache, Alexandru and Voinescu, Andrei and Moore, Andrew W and Antichi, Gianni and W{\'o}jcik, Marcin},
  booktitle={Proceedings of the Conference of the ACM Special Interest Group on Data Communication},
  pages={29--42},
  year={2017}
}

@inproceedings{eqds,
  title={An edge-queued datagram service for all datacenter traffic},
  author={Olteanu, Vladimir and Eran, Haggai and Dumitrescu, Dragos and Popa, Adrian and Baciu, Cristi and Silberstein, Mark and Nikolaidis, Georgios and Handley, Mark and Raiciu, Costin},
  booktitle={19th USENIX Symposium on Networked Systems Design and Implementation (NSDI 22)},
  pages={761--777},
  year={2022}
}

@inproceedings{swift,
  title={Swift: Delay is simple and effective for congestion control in the datacenter},
  author={Kumar, Gautam and Dukkipati, Nandita and Jang, Keon and Wassel, Hassan MG and Wu, Xian and Montazeri, Behnam and Wang, Yaogong and Springborn, Kevin and Alfeld, Christopher and Ryan, Michael and others},
  booktitle={Proceedings of the Annual conference of the ACM Special Interest Group on Data Communication on the applications, technologies, architectures, and protocols for computer communication},
  pages={514--528},
  year={2020}
}

@inproceedings{plb,
  title={PLB: congestion signals are simple and effective for network load balancing},
  author={Qureshi, Mubashir Adnan and Cheng, Yuchung and Yin, Qianwen and Fu, Qiaobin and Kumar, Gautam and Moshref, Masoud and Yan, Junhua and Jacobson, Van and Wetherall, David and Kabbani, Abdul},
  booktitle={Proceedings of the ACM SIGCOMM 2022 Conference},
  pages={207--218},
  year={2022}
}

@inproceedings{dcpim,
  title={dcPIM: Near-optimal proactive datacenter transport},
  author={Cai, Qizhe and Arashloo, Mina Tahmasbi and Agarwal, Rachit},
  booktitle={Proceedings of the ACM SIGCOMM 2022 Conference},
  pages={53--65},
  year={2022}
}

@inproceedings{fastpass,
  title={Fastpass: A centralized" zero-queue" datacenter network},
  author={Perry, Jonathan and Ousterhout, Amy and Balakrishnan, Hari and Shah, Devavrat and Fugal, Hans},
  booktitle={Proceedings of the 2014 ACM conference on SIGCOMM},
  pages={307--318},
  year={2014}
}

@article{pfabric,
  title={pFabric: Minimal near-optimal datacenter transport},
  author={Alizadeh, Mohammad and Yang, Shuang and Sharif, Milad and Katti, Sachin and McKeown, Nick and Prabhakar, Balaji and Shenker, Scott},
  journal={ACM SIGCOMM Computer Communication Review},
  volume={43},
  number={4},
  pages={435--446},
  year={2013},
  publisher={ACM New York, NY, USA}
}

@inproceedings{clove,
  title={Clove: Congestion-aware load balancing at the virtual edge},
  author={Katta, Naga and Ghag, Aditi and Hira, Mukesh and Keslassy, Isaac and Bergman, Aran and Kim, Changhoon and Rexford, Jennifer},
  booktitle={Proceedings of the 13th International Conference on emerging Networking EXperiments and Technologies},
  pages={323--335},
  year={2017}
}

@inproceedings{conga,
  title={CONGA: Distributed congestion-aware load balancing for datacenters},
  author={Alizadeh, Mohammad and Edsall, Tom and Dharmapurikar, Sarang and Vaidyanathan, Ramanan and Chu, Kevin and Fingerhut, Andy and Lam, Vinh The and Matus, Francis and Pan, Rong and Yadav, Navindra and others},
  booktitle={Proceedings of the 2014 ACM conference on SIGCOMM},
  pages={503--514},
  year={2014}
}

@inproceedings{letitflowlet,
  title={Let it flow: Resilient asymmetric load balancing with flowlet switching},
  author={Vanini, Erico and Pan, Rong and Alizadeh, Mohammad and Taheri, Parvin and Edsall, Tom},
  booktitle={14th USENIX Symposium on Networked Systems Design and Implementation (NSDI 17)},
  pages={407--420},
  year={2017}
}

@article{kandulaflowlet,
  title={Dynamic load balancing without packet reordering},
  author={Kandula, Srikanth and Katabi, Dina and Sinha, Shantanu and Berger, Arthur},
  journal={ACM SIGCOMM Computer Communication Review},
  volume={37},
  number={2},
  pages={51--62},
  year={2007},
  publisher={ACM New York, NY, USA}
}

@inproceedings{meta,
  title={{RDMA} over {Ethernet} for distributed training at {MEta} scale},
  author={Gangidi, Adithya and Miao, Rui and Zheng, Shengbao and Bondu, Sai Jayesh and Goes, Guilherme and Morsy, Hany and Puri, Rohit and Riftadi, Mohammad and Shetty, Ashmitha Jeevaraj and Yang, Jingyi and others},
  booktitle={Proceedings of the ACM SIGCOMM 2024 Conference},
  pages={57--70},
  year={2024}
}

@inproceedings{alibaba,
  title={Alibaba hpn: A data center network for large language model training},
  author={Qian, Kun and Xi, Yongqing and Cao, Jiamin and Gao, Jiaqi and Xu, Yichi and Guan, Yu and Fu, Binzhang and Shi, Xuemei and Zhu, Fangbo and Miao, Rui and others},
  booktitle={Proceedings of the ACM SIGCOMM 2024 Conference},
  pages={691--706},
  year={2024}
}

@misc{h100,
    title={NVIDIA H100 Tensor Core GPU},
    howpublished={\url{https://www.nvidia.com/en-us/data-center/h100/}},
    author={NVIDIA Corporation},
}

@misc{spectrumx,
    title={NVIDIA Spectrum SN5000 Series Switches},
    howpublished={\url{https://nvdam.widen.net/s/mmvbnpk8qk/networking-ethernet-switches-sn5000-datasheet-us}},
    author={{NVIDIA Corporation}},
}

@misc{broadcomtom,
    title={Tomahawk3 / BCM56980 Series},
    howpublished={\url{https://www.broadcom.com/products/ethernet-connectivity/switching/strataxgs/bcm56980-series}},
    author={{Broadcom}} 
}

@misc{csghtsim,
    title={csg-htsim},
    howpublished={\url{https://github.com/Broadcom/csg-htsim}},
    year={2023} 
}

@inproceedings{mptcp,
  title={Opportunistic mobility with multipath TCP},
  author={Raiciu, Costin and Niculescu, Dragos and Bagnulo, Marcelo and Handley, Mark James},
  booktitle={Proceedings of the sixth international workshop on MobiArch},
  pages={7--12},
  year={2011}
}

@misc{raptorq-rfc,
    series =    {Request for Comments},
    number =    6330,
    howpublished =  {RFC 6330},
    publisher = {RFC Editor},
    doi =       {10.17487/RFC6330},
    url =       {https://www.rfc-editor.org/info/rfc6330},
    author =    {Lorenz Minder and Amin Shokrollahi and Mark Watson and Michael Luby and Thomas Stockhammer},
    title =     {{RaptorQ Forward Error Correction Scheme for Object Delivery}},
    pagetotal = 69,
    year =      2011,
    month =     aug,
}

@inproceedings{flowbender,
  title={Flowbender: Flow-level adaptive routing for improved latency and throughput in datacenter networks},
  author={Kabbani, Abdul and Vamanan, Balajee and Hasan, Jahangir and Duchene, Fabien},
  booktitle={Proceedings of the 10th ACM International on Conference on emerging Networking Experiments and Technologies},
  pages={149--160},
  year={2014}
}

@inproceedings{detail,
  title={DeTail: Reducing the flow completion time tail in datacenter networks},
  author={Zats, David and Das, Tathagata and Mohan, Prashanth and Borthakur, Dhruba and Katz, Randy},
  booktitle={Proceedings of the ACM SIGCOMM 2012 conference on Applications, technologies, architectures, and protocols for computer communication},
  pages={139--150},
  year={2012}
}

@article{presto,
  title={Presto: Edge-based load balancing for fast datacenter networks},
  author={He, Keqiang and Rozner, Eric and Agarwal, Kanak and Felter, Wes and Carter, John and Akella, Aditya},
  journal={ACM SIGCOMM Computer Communication Review},
  volume={45},
  number={4},
  pages={465--478},
  year={2015},
  publisher={ACM New York, NY, USA}
}

@inproceedings{rps,
  title={On the impact of packet spraying in data center networks},
  author={Dixit, Advait and Prakash, Pawan and Hu, Y Charlie and Kompella, Ramana Rao},
  booktitle={2013 Proceedings IEEE INFOCOM},
  pages={2130--2138},
  year={2013},
  organization={IEEE}
}

@inproceedings{hula,
  title={Hula: Scalable load balancing using programmable data planes},
  author={Katta, Naga and Hira, Mukesh and Kim, Changhoon and Sivaraman, Anirudh and Rexford, Jennifer},
  booktitle={Proceedings of the Symposium on SDN Research},
  pages={1--12},
  year={2016}
}

@inproceedings{expeditus,
  title={Expeditus: Congestion-aware load balancing in clos data center networks},
  author={Wang, Peng and Xu, Hong and Niu, Zhixiong and Han, Dongsu and Xiong, Yongqiang},
  booktitle={Proceedings of the Seventh ACM Symposium on Cloud Computing},
  pages={442--455},
  year={2016}
}

@inproceedings{hermes,
  title={Resilient datacenter load balancing in the wild},
  author={Zhang, Hong and Zhang, Junxue and Bai, Wei and Chen, Kai and Chowdhury, Mosharaf},
  booktitle={Proceedings of the Conference of the ACM Special Interest Group on Data Communication},
  pages={253--266},
  year={2017}
}

@inproceedings{hedera,
  title={Hedera: dynamic flow scheduling for data center networks.},
  author={Al-Fares, Mohammad and Radhakrishnan, Sivasankar and Raghavan, Barath and Huang, Nelson and Vahdat, Amin and others},
  booktitle={Nsdi},
  volume={10},
  number={8},
  pages={89--92},
  year={2010},
  organization={San Jose, USA}
}

@inproceedings{mahout,
  title={Mahout: Low-overhead datacenter traffic management using end-host-based elephant detection},
  author={Curtis, Andrew R and Kim, Wonho and Yalagandula, Praveen},
  booktitle={2011 Proceedings IEEE INFOCOM},
  pages={1629--1637},
  year={2011},
  organization={IEEE}
}

@inproceedings{microte,
  title={MicroTE: Fine grained traffic engineering for data centers},
  author={Benson, Theophilus and Anand, Ashok and Akella, Aditya and Zhang, Ming},
  booktitle={Proceedings of the seventh conference on emerging networking experiments and technologies},
  pages={1--12},
  year={2011}
}

@inproceedings{localflow,
  title={Scalable, optimal flow routing in datacenters via local link balancing},
  author={Sen, Siddhartha and Shue, David and Ihm, Sunghwan and Freedman, Michael J},
  booktitle={Proceedings of the ninth ACM conference on Emerging networking experiments and technologies},
  pages={151--162},
  year={2013}
}

@misc{ecmprfc,
    series =    {Request for Comments},
    number =    2992,
    howpublished =  {RFC 2992},
    publisher = {RFC Editor},
    doi =       {10.17487/RFC2992},
    url =       {https://www.rfc-editor.org/info/rfc2992},
    author =    {Christian Hopps},
    title =     {{Analysis of an Equal-Cost Multi-Path Algorithm}},
    pagetotal = 8,
    year =      2000,
    month =     nov,
}

@misc{ethereal,
      title={Ethereal: Divide and Conquer Network Load Balancing in Large-Scale Distributed Training}, 
      author={Vamsi Addanki and Prateesh Goyal and Ilias Marinos and Stefan Schmid},
      year={2025},
      eprint={2407.00550},
      archivePrefix={arXiv},
      primaryClass={cs.NI},
      url={https://arxiv.org/abs/2407.00550}, 
}

@misc{falcon,
    title={Google opens Falcon, a reliable low-latency hardware transport, to the ecosystem},
    author={Dan Lenoski and Nandita Dukkipati},
    howpublished={\url{https://cloud.google.com/blog/topics/systems/introducing-falcon-a-reliable-low-latency-hardware-transport}},
    year={2023}
}

@misc{uec,
    title={Ultra Ethernet Consortium},
    howpublished={\url{https://ultraethernet.org/}},
    year={2026}
}

@misc{uec-spec,
    title={Ultra Ethernet Specification v1.0.2},
    author={Ultra Ethernet Consortium},
    day={28},
    month={Jan},
    year={2026},
    howpublished={\url{https://ultraethernet.org/wp-content/uploads/sites/20/2026/01/UE-Specification-1.0.2-1.pdf}}
}

@misc{nvidia-ooo,
    title={Out-of-order Data Placement},
    author={NVIDIA},
    howpublished={\url{https://docs.nvidia.com/doca/archive/2-9-4/out-of-order-data-placement/index.html}}
}

@INPROCEEDINGS{dragonfly-ar,
  author={Shpiner, Alexander and Haramaty, Zachy and Eliad, Saar and Zdornov, Vladimir and Gafni, Barak and Zahavi, Eitan},
  booktitle={2017 IEEE 3rd International Workshop on High-Performance Interconnection Networks in the Exascale and Big-Data Era (HiPINEB)}, 
  title={Dragonfly+: Low Cost Topology for Scaling Datacenters}, 
  year={2017},
  volume={},
  number={},
  pages={1-8},
  doi={10.1109/HiPINEB.2017.11}}

@inproceedings{conweave,
author = {Song, Cha Hwan and Khooi, Xin Zhe and Joshi, Raj and Choi, Inho and Li, Jialin and Chan, Mun Choon},
title = {Network Load Balancing with In-network Reordering Support for RDMA},
year = {2023},
isbn = {9798400702365},
publisher = {Association for Computing Machinery},
address = {New York, NY, USA},
url = {https://doi.org/10.1145/3603269.3604849},
doi = {10.1145/3603269.3604849},
booktitle = {Proceedings of the ACM SIGCOMM 2023 Conference},
pages = {816–831},
numpages = {16},
location = {New York, NY, USA},
series = {ACM SIGCOMM '23}
}

@INPROCEEDINGS{rocesack,
  author={Nie, Yang and Shi, Zheng and Chen, Xinyi and Qian, Liguo},
  booktitle={2022 IEEE 6th Advanced Information Technology, Electronic and Automation Control Conference (IAEAC )}, 
  title={An Out-of-Order Packet Processing Algorithm of RoCE Based on Improved SACK}, 
  year={2022},
  volume={},
  number={},
  pages={1402-1408},
  doi={10.1109/IAEAC54830.2022.9929858}}

@inproceedings{bonatoreps,
  title={{REPS}: Recycled Entropy Packet Spraying for Adaptive Load Balancing and Failure Mitigation},
  author={Tommaso Bonato and Abdul Kabbani and Ahmad Ghalayini and Michael Papamichael and Mohammad Dohadwala and Lukas Gianinazzi and Mikhail Khalilov and Elias Achermann and Daniele De Sensi and Torsten Hoefler},
  booktitle={{EuroSys}},
  year={2026}
}

@ARTICLE{raptor-theory,
  author={Shokrollahi, A.},
  journal={IEEE Transactions on Information Theory}, 
  title={Raptor codes}, 
  year={2006},
  volume={52},
  number={6},
  pages={2551-2567},
  doi={10.1109/TIT.2006.874390}}

@INPROCEEDINGS {ltcodes,
author = { Luby, Michael },
booktitle = { 2013 IEEE 54th Annual Symposium on Foundations of Computer Science },
title = {{ LT Codes }},
year = {2002},
volume = {},
ISSN = {0272-5428},
pages = {271},
doi = {10.1109/SFCS.2002.1181950},
url = {https://doi.ieeecomputersociety.org/10.1109/SFCS.2002.1181950},
publisher = {IEEE Computer Society},
address = {Los Alamitos, CA, USA},
month =Nov}

@InProceedings{ring-mpich,
author="Thakur, Rajeev
and Gropp, William D.",
editor="Dongarra, Jack
and Laforenza, Domenico
and Orlando, Salvatore",
title="Improving the Performance of Collective Operations in MPICH",
booktitle="Recent Advances in Parallel Virtual Machine and Message Passing Interface",
year="2003",
publisher="Springer Berlin Heidelberg",
address="Berlin, Heidelberg",
pages="257--267",
isbn="978-3-540-39924-7"
}

@misc{parallelism,
    title={Paradigms of Parallelism},
    howpublished={\url{https://colossalai.org/docs/concepts/paradigms_of_parallelism/}},
    author={Shenggui Li and Siqi Mai}
}

@misc{ncclcollectives,
    title={Collective Operations},
    author={NVIDIA Corporation},
    howpublished={\url{https://docs.nvidia.com/deeplearning/nccl/user-guide/docs/usage/collectives.html}},
    year={2020}
}

@misc{nccl,
    title={NVIDIA Collective Communications Library (NCCL)},
    author={NVIDIA Corporation},
    howpublished={\url{https://developer.nvidia.com/nccl}},
    year={2025}
}

@misc{nccl-ring,
    title={{NCCL}: Accelerated multi-{GPU} collective communications},
    author={Cliff Woolley},
    howpublished={\url{https://images.nvidia.com/events/sc15/pdfs/NCCL-Woolley.pdf}}
}

@inproceedings{sccl,
author = {Cai, Zixian and Liu, Zhengyang and Maleki, Saeed and Musuvathi, Madanlal and Mytkowicz, Todd and Nelson, Jacob and Saarikivi, Olli},
title = {Synthesizing optimal collective algorithms},
year = {2021},
isbn = {9781450382946},
publisher = {Association for Computing Machinery},
address = {New York, NY, USA},
url = {https://doi.org/10.1145/3437801.3441620},
doi = {10.1145/3437801.3441620},
booktitle = {Proceedings of the 26th ACM SIGPLAN Symposium on Principles and Practice of Parallel Programming},
pages = {62–75},
numpages = {14},
location = {Virtual Event, Republic of Korea},
series = {PPoPP '21}
}

@misc{openai-ms-blog,
    title={Microsoft announces new supercomputer, lays out vision for future AI work},
    howpublished={\url{https://news.microsoft.com/source/features/ai/openai-azure-supercomputer/}},
    author={Jennifer Langston},
    year={2020}
}

@misc{strack,
      title={STrack: A Reliable Multipath Transport for AI/ML Clusters}, 
      author={Yanfang Le and Rong Pan and Peter Newman and Jeremias Blendin and Abdul Kabbani and Vipin Jain and Raghava Sivaramu and Francis Matus},
      year={2024},
      eprint={2407.15266},
      archivePrefix={arXiv},
      primaryClass={cs.NI},
      url={https://arxiv.org/abs/2407.15266}, 
}

@inproceedings{bytedance-glb,
author = {Qi, Chenchen and Wu, Wenfei and Wang, Yongcan and He, Keqiang and Kao, Yu-Hsiang (Sean) and He, Zongying and Yen, Chen-Yu and Jiang, Zhuo and Luo, Feng and Anubolu, Surendra and Gao, Yanjin and Lin, Bingfeng and Ni, Wenda and Yang, Yiming and Wei, Donglin and Zhou, Boyang and Wang, Jian and Ding, Shan},
title = {SGLB: Scalable and Robust Global Load Balancing in Commodity AI Clusters},
year = {2025},
isbn = {9798400715242},
publisher = {Association for Computing Machinery},
address = {New York, NY, USA},
url = {https://doi.org/10.1145/3718958.3750527},
doi = {10.1145/3718958.3750527},
booktitle = {Proceedings of the ACM SIGCOMM 2025 Conference},
pages = {626–644},
numpages = {19},
location = {S\~{a}o Francisco Convent, Coimbra, Portugal},
series = {SIGCOMM '25}
}

@inproceedings{alibaba-stellar,
author = {Lu, Jie and Gao, Jiaqi and Feng, Fei and He, Zhiqiang and Zheng, Menglei and Liu, Kun and He, Jun and Liao, Binbin and Xu, Suwei and Sun, Ke and Mo, Yongjia and Peng, Qinghua and Luo, Jilie and Li, Qingxu and Lu, Gang and Wang, Zishu and Dong, Jianbo and He, Kunling and Cheng, Sheng and Cao, Jiamin and Jiao, Hairong and Zhang, Pengcheng and Ma, Shu and Zhu, Lingjun and Shi, Chao and Zhang, Yangming and Chen, Yiquan and Wang, Wei and Zhu, Shuhong and Li, Xingru and Wang, Qiang and Liu, Jiang and Wang, Chao and Lin, Wei and Zhai, Ennan and Wu, Jiesheng and Liu, Qiang and Fu, Binzhang and Cai, Dennis},
title = {Alibaba Stellar: A New Generation RDMA Network for Cloud AI},
year = {2025},
isbn = {9798400715242},
publisher = {Association for Computing Machinery},
address = {New York, NY, USA},
url = {https://doi.org/10.1145/3718958.3750539},
doi = {10.1145/3718958.3750539},
booktitle = {Proceedings of the ACM SIGCOMM 2025 Conference},
pages = {453–466},
numpages = {14},
location = {S\~{a}o Francisco Convent, Coimbra, Portugal},
series = {SIGCOMM '25}
}

@misc{sack,
    series =    {Request for Comments},
    number =    2018,
    howpublished =  {RFC 2018},
    publisher = {RFC Editor},
    doi =       {10.17487/RFC2018},
    url =       {https://www.rfc-editor.org/info/rfc2018},
    author =    {Sally Floyd and Jamshid Mahdavi and Matt Mathis and Dr. Allyn Romanow},
    title =     {{TCP Selective Acknowledgment Options}},
    pagetotal = 12,
    year =      1996,
    month =     oct,
}

@misc{wang2024auxiliarylossfreeloadbalancingstrategy,
      title={Auxiliary-Loss-Free Load Balancing Strategy for Mixture-of-Experts}, 
      author={Lean Wang and Huazuo Gao and Chenggang Zhao and Xu Sun and Damai Dai},
      year={2024},
      eprint={2408.15664},
      archivePrefix={arXiv},
      primaryClass={cs.LG},
      url={https://arxiv.org/abs/2408.15664}, 
}

@misc{bgp-in-dcs-rfc,
    series =    {Request for Comments},
    number =    7938,
    howpublished =  {RFC 7938},
    publisher = {RFC Editor},
    doi =       {10.17487/RFC7938},
    url =       {https://www.rfc-editor.org/info/rfc7938},
    author =    {Petr Lapukhov and Ariff Premji and Jon Mitchell},
    title =     {{Use of BGP for Routing in Large-Scale Data Centers}},
    pagetotal = 35,
    year =      2016,
    month =     aug,
}

@inproceedings {bgp-in-dcs,
author = {Anubhavnidhi Abhashkumar and Kausik Subramanian and Alexey Andreyev and Hyojeong Kim and Nanda Kishore Salem and Jingyi Yang and Petr Lapukhov and Aditya Akella and Hongyi Zeng},
title = {Running {BGP} in Data Centers at Scale},
booktitle = {18th USENIX Symposium on Networked Systems Design and Implementation (NSDI 21)},
year = {2021},
isbn = {978-1-939133-21-2},
pages = {65--81},
url = {https://www.usenix.org/conference/nsdi21/presentation/abhashkumar},
publisher = {USENIX Association},
month = apr
}

@misc{deepspeed-moe,
      title={DeepSpeed-MoE: Advancing Mixture-of-Experts Inference and Training to Power Next-Generation AI Scale}, 
      author={Samyam Rajbhandari and Conglong Li and Zhewei Yao and Minjia Zhang and Reza Yazdani Aminabadi and Ammar Ahmad Awan and Jeff Rasley and Yuxiong He},
      year={2022},
      eprint={2201.05596},
      archivePrefix={arXiv},
      primaryClass={cs.LG},
      url={https://arxiv.org/abs/2201.05596}, 
}

\clearpage
\appendix{}

\section{Sending Rate for Randomized Failures}
\label{app:failure}

As mentioned in \S\ref{sec:lb}, we randomly fail links with a configured probability. 
Given the random process, we cannot ensure that there is sufficient capacity in the network to serve the demand.
Further, even if there is sufficient capacity for a given rate (even line rate), it is not clear that any load balancing scheme will be able to manage the demand.

With failures, there may be a possible routing of flows that can serve the traffic at some optimal rate $\rho_{opt}$. However, this may only be achievable with some traffic engineering approach that can route flows according to the solution of a multi-commodity maximum concurrent flow problem (MCFP) -- \ie simply load balancing across available paths will not reach the solution that allows $\rho_{opt}$. For example, some flows may only be able to traverse one path, and therefore other flows should specifically not use that path.

Thus, we instead use a more realistic model of what load-balancing schemes do: attempt to spread load equally across all available paths.
Accordingly, we define $\rhomax$ as the maximum possible rate the network can serve under the assumption that all flows will be split evenly among their available paths.
The value of $\rhomax$ can be rather easily calculated by counting the units of flow on each link for each flow and finding the maximally loaded link in the network.
Assume that this bottleneck link has total flow $F$ on it and every link in the network has bandwidth $B$. 
Then the per-flow rate, $\rhomax = \frac{B}{F}$.

We note that this is a rather pessimistic rate (there may be wasted capacity in the network), but it ensures that the demand will not congest the network with failures. Further, we find that despite being a lower bound, different load-balancing schemes are still differentiated by it.

\section{Permutation CCT Lower-Bound}\label{sec:lowerbound}

\begin{figure}
    \centering
    \includegraphics[width=0.65\linewidth]{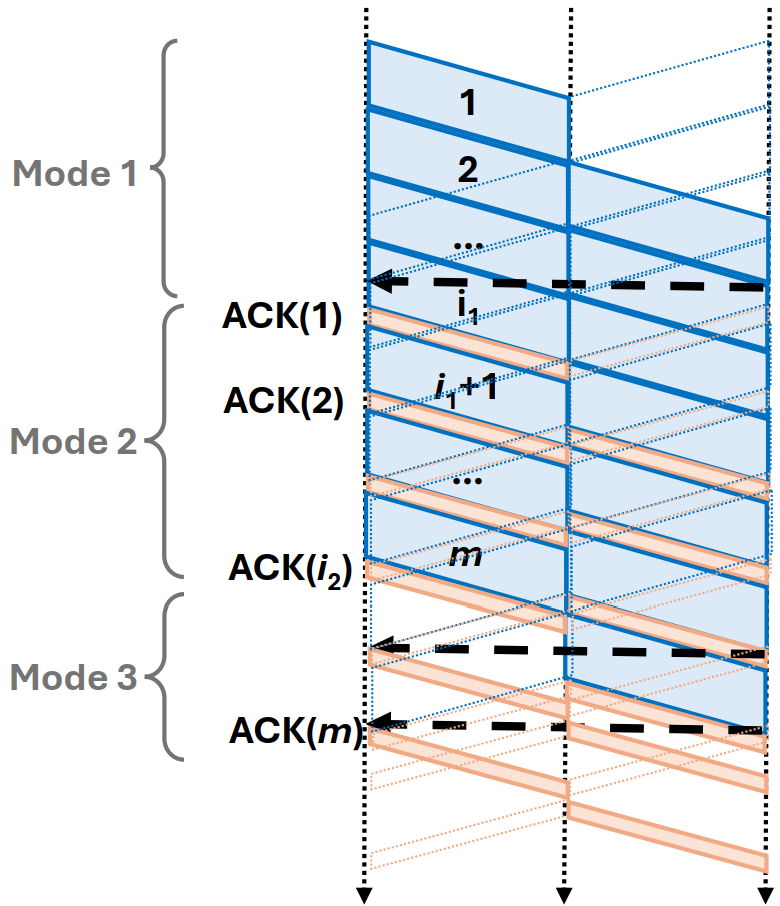}
    \caption{Illustration of the symmetric dynamics of the permutation collective. 
    }
    \label{fig:ACK}
    \vspace{-1em}
\end{figure}

\mypar{Outline} \Cref{fig:ACK} illustrates how permutation collectives in fat trees exhibit interesting interactions between data and ACK packets, as each host is at the same time a sender and a receiver. When computing the CCT, we find that there are in fact three successive modes of communication: (1)~Data packets only, then (2)~interleaved data and ACK packets, and finally (3)~ACK packets only. In order to provide a lower bound for the CCT, we need to take these three modes into account. 

\mypar{Assumptions} In the remainder, we focus on a node that receives a flow of data packets going through $\H=6$ hops, and sends another flow of ACK packets traversing $\H$ hops as well. Let $\Td$, $\Ta$, $\Tg$ and $\p$ respectively denote the data and ACK packet serialization times, the gap time between packets, and the one-way propagation time. Also, let $\d=\Td+\Tg$ and $\a=\Ta+\Tg$.

\mypar{Mode 1: only data} The last bit of the first data packet of each message goes through at least a propagation time of $\p$ and $\H$ serialization delays of $\Td$, thus packet 1 arrives at time $t_1=\p+\H\cdot\Td$.

When it gets to the receiver, this receiver is in the middle of sending data packet indexed $i_1$. It needs $\Td+(i_1-1)\cdot \d$ to send these $i_1$ data packets. Therefore, putting both equations together yields $i_1$:

\begin{equation}
    i_1 = \ceil{\frac{\p+(\H-1)\cdot\Td}{\d}}+1
\end{equation}

\mypar{Example} With the paper's settings, $i_1=78$ data packets. In other words, each sender sends at least 78 consecutive data packets before it  sends ACKs. OF course, if $\m<i_1$, then it only sends $\m$ data packets.

\mypar{Mode 2: interleaved data and ACK packets} Once the host starts interleaving data and ACK packets, we assume that the host policy performs round-robin between the two, as long as there is data left. 

Let's study when the first ACK arrives. Before, we saw that $i_1$ data packets are sent. We also saw that packet 1 arrives at time $t_1=\p+\H\cdot\Td$. Therefore, packet $i_1$ arrives at time $t_{i_1}=T_1+(i_1-1) \d$. The first ACK for packet 1 arrives just after, at $t^{\text{ACK}}_1=T_{i_1}+\a$.

Afterwards, all packets are sent in an interleaved way, data-ACK-data-ACK-etc. Therefore, the ACK for packet $i$ arrives at $t^{\text{ACK}}_i=t^{\text{ACK}}_1+(i-1)(\d+\a)$. 

\mypar{Mode 3: only ACKs} At the end, the hosts send the $\m\th$ message, and don't need to interleave data and ACKs anymore. This is a symmetric case from $i_1$: it happens as they send ACK $i_2$, with $i_2=\m-i_1+1$. For example, with $\m=256$ and $i_1=78$, it happens at the ACK for packet $179$.

Beyond index $i_2$, we have two constraints. First, ACKs should have time to be transmitted. We get a first lower-bound time $t^{\text{ACK}}_i(1)=t^{\text{ACK}}_{i_2}+(i-i_2)(\a)$. Second, ACK packets can obviously only be sent if they acknowledge a data packet that has arrived. Just like packet $i_1$ arrived just before ACK $1$, so does packet $i$ arrive just before ACK $i-(i_1-1)$. Thus, we obtain a second lower-bound time: $t^{\text{ACK}}_i(2)=t^{\text{ACK}}_{i-(i_1-1)}+(H-1)(\a)+\p$. Finally, the lower bound is the maximum of both constraints: for $i_2+1 \leq i \leq \m$,$
t^{\text{ACK}}_i=
\max\para{t^{\text{ACK}}_i(1),
     t^{\text{ACK}}_i(2)}
$.

\mypar{Example} With our default assumptions, we get a lower bound of $17.05694\us$ for a message size $\m=256.$ We want to compare that to smallest value that we can get. We only send a flow from the first host to the last and back in a small network ($\k=4$), using all algorithms, iterate 1000 times, and take the minimum over all possibilities, we measure in practice a value of $17.0587\us$ (with $\k=8$, we get $17.0609\us$, which is less tight). That is, the difference is $10^{-4}$ of the CCT, less than a packet time, showing the tightness of the lower bound.

\section{Proof of \cref{thm:sync}: Synchronization} \label{sec:sync}

\subsection{Proof Overview}
We now prove the synchronization properties of \simplerr and \jsq. The synchronization derives from three properties:

\mypar{1. Perfect match for first packets} Consider an edge switch without internal traffic.
Assume that flow $1$ sends the first packet, which is forwarded to uplink port $i$. Each of the
other flows then sends one packet that is forwarded to a distinct uplink port: (i)~In RR, because of the round-robin rotation; and (ii)~in AR, because the queues of these ports are initially empty, then any port that is assigned a packet gets a queue size of 1 and is removed from consideration for a duration $T$. At the end, there is a perfect 1-to-1 match of packets to uplink ports.

\mypar{2. Sticky flows} When the second packet of flow $1$ arrives after a period $T$: (i)~in $RR$, it will be sent again to port $i$ due to the round-robin periodic rotation, and (ii)~in $AR$, the first packet of flow 1 will have just left, since it takes $\pkt/\C<T$ to service it. Thus port $i$ is the only empty port and the second picks it. In both cases, port $i$ is a sticky port for flow 1. Likewise, each other flow will have its own sticky port.

\mypar{3. Aggregation switch layout}
Now we understand how flows can be stuck onto specific edge-aggregation links. In addition, in a fat tree, a flow stuck to aggregation switch $i$ of a source pod will also be stuck to aggregation switch $i$ of a destination pod (black and red circles in \cref{fig:failure-diagram}). Thus, two flows from different source pods but destined to the same edge switch may have picked the same aggregation switch $i$, leading to contention at the link from the destination aggregation switch to the destination edge switch and to linear growth of its queue.

\mypar{Proof outline} We focus on the synchronization of (at least) two flows that end up sharing a single aggregation-edge link.  When this happens, the queue growth is linear. The probability of such a synchronization yields  a lower bound on the general synchronization probability. We show that as switches becomes larger, this probability goes to 1.

\subsection{Proof}
\begin{proof}[Proof of \cref{thm:sync}]

More specifically, we show that \simplerr exhibits such a core-aggregation link synchronization whenever at least two flows satisfy three specific properties; and \jsq exhibits it whenever four properties are satisfied. We will first list the four properties, then show that the general probability goes to 1 for larger switches.

\mypar{Property 1: Red flows} For the sake of discussion, we will denote \textit{red flows} as flows originating at a \textit{red edge switch}, which is an edge switch that satisfies two properties: (i)~It has no internal traffic, \ie all its flows are northbound; and (ii)~not all its flows are destined to the exact same destination edge switch. The goal of the first condition is to ensure that the number of northbound flows equals the number of northbound ports, which facilitates synchronization. The goal of the second condition is to avoid a rare case where all of the flows to the same destination effectively come from source aggregation switches with different pod indices, and therefore from distinct destination aggregation switches because of the fat-tree properties, which prevents the synchronization.

The probability for the first condition that all traffic is northbound is the probability that the first flow is northbound, times the probability for the second flow, \etc 
\begin{align}
    P_{\text{northbound}} &= 
    \frac{\n-\k/2}{\n-1} \cdot \frac{\n-\k/2-1}{\n-2} \cdot \cdots \cdot  \frac{\n-\k+1}{\n-\k/2} \\
    &= \prod_{i=0}^{\k/2-1} \frac{\n-\k/2-i}{\n-1-i} \label{eq:northbound}
\end{align}

The probability for the second condition that all traffic targets the same outside hotspot edge switch is the product of the probability that the first host picks any such edge, 
times the probability that other hosts follow:
\begin{equation}
    P_{\text{hotspot}} = {\frac{\n-\k/2}{\n-1}} \cdot {\prod_{i=1}^{\k/2-1} \frac{\k/2-i}{\n-1-i}}
\end{equation}
Let $A$ denote the event that all flows are northbound, and $B$ that all flows target the exact same outside hotspot. An  edge switch is red (event $C$) if $A$ happens but $B$ doesn't. So $P(C)=P(A)-P(A \cap B) = P(A)-P(B),$ thus
\begin{equation}
    P_{Red} = P_{\text{northbound}} - P_{\text{hotspot}}.
\end{equation}

\mypar{Property 2: Safe flows} To prove synchronization in \jsq aggregation-edge links, we require that when a packet needs to select an uplink port, there exists \textit{exactly one port with the strictly shortest queue}. Furthermore, this specific port must be identical to the one used by the previous packet of the same flow.

To ensure this continuity, we focus on a flow, and define strict timing constraints regarding the service completion of competing flows relative to the service completion of our flow's previous packet:
\begin{enumerate}
    \item[(i)] No packets in other queues completed transmission within the the inter-arrival gap duration $T_{IPG}$  \textit{before} the service completion of the flow's packet. The reason is that if a packet from a competing flow were to finish within this window, its queue would become empty. When the next packet of that competing flow arrives, it will observe two empty queues (its own and the target flow's). Consequently, it may select the target flow's port with equal probability, which we want to avoid.
    
    \item[(ii)] No packets in other queues completed transmission within the duration $T_{IPG}$ \textit{after} the service completion of the flow's packet. The reason is that if a competing flow finishes within this window, its queue also becomes empty. The next packet of our target flow arrives exactly $T_{IPG}$ after the service completion of the last packet from our considered flow. Thus, it will encounter two empty queues (its original port and the competitor's). It may therefore select the competitor's port instead of preserving its path.
\end{enumerate}
Combined, these conditions imply that no other flows initiated their transmission time within the critical window of $T_{IPG}$ before and after the requested flow.

\textbf{Explanation:} Since no other packets completed service during the critical window, we encounter a state where there are exactly $\k/2 - 1$ packets distributed across the $\k/2$ uplink queues. Under the \jsq policy, this implies that every queue holds exactly one packet, except for a single empty queue.
We know that no state change has occurred since the previous packet from our flow departed this specific queue. Therefore, the AR algorithm will deterministically select this exact same port again. This mechanism ensures that all packets from a specific flow are directed to the same Aggregation switch, effectively mimicking static routing.

\textbf{Probability Calculations:}
We define the total packet cycle time as $T_{SERVICE} = T_{DATA} + T_{IPG}$.
First, consider the interaction between our flow and a \textit{single} neighboring host. The probability that this neighbor transmits a packet that temporally overlaps with the critical inter-arrival gap ($T_{IPG}$) is proportional to the ratio of the vulnerable window size to the total cycle time:
\begin{equation}
    P_{\text{single\_overlap}} = \frac{2 \cdot T_{IPG}}{T_{SERVICE}}
\end{equation}

Therefore, the probability that this specific neighbor remains "silent" during the critical window (does not interfere) is the complement:
\begin{equation}
    P_{\text{safe\_pair}} = 1 - P_{\text{single\_overlap}} = 1 - \frac{2 \cdot T_{IPG}}{T_{SERVICE}}
\end{equation}

The edge switch connects $m = \frac{k}{2} $ hosts. For our flow to be safe, it must not overlap with \textit{any} of the other $\frac{k}{2} - 1$ hosts connected to the same leaf. therefore:
\begin{equation}
        P_{\text{safe}} = \left( 1 - \frac{2 \cdot T_{IPG}}{T_{SERVICE}} \right)^{\frac{\k}{2} - 1}
\end{equation}

\mypar{Property 3: Same aggregation switch index}
In a Fat-Tree topology, the routing structure is symmetric. If a packet ascends from the $i$-th Aggregation switch in the source pod to the Core layer, it is structurally forced to descend to the $i$-th Aggregation switch in the destination pod.

Therefore, for two flows to converge at the same specific Destination Aggregation switch (and subsequently compete for the downlink), they must select the exact same Aggregation switch index when traffic ascends from the Edge switch to the Aggregation layer.

Since there are $\k/2$ available aggregation switches for the upstream path, the probability that two independent flows select the same index is:
\begin{equation}
    P_{\text{same agg.}} = \frac{1}{\k/2} = \frac{2}{K}
\end{equation}

\mypar{Property 4: Same destination edge switch}
Both flows are destined for different hosts that reside under the \textit{same Destination edge switch}.

\mypar{Probability calculation (given a red flow)}
We calculate the ratio between the valid targets in the specific destination leaf and the total pool of valid targets outside the source leaf:
\begin{itemize}
    \item \textbf{Numerator:} The number of potential hosts in the specific target leaf (excluding the one already chosen by the first flow).
    \item \textbf{Denominator:} The total number of valid destination hosts in the network, excluding all hosts in the source leaf (since the flow is Red) and the host taken by the first flow.
\end{itemize}

\begin{equation}
    P_{\text{same dest edge}} = \frac{{\frac{\k}{2}-1}^{}}{{n - 1 - \frac{\k}{2}}_{}} = \frac{\frac{\k}{2}-1}{\frac{\k^3}{4}-1-\frac{\k}{2}}
\end{equation}
\mypar{Total collision probability}
A collision occurs when two hosts meet all four categories simultaneously.

\begin{equation}
    P_{\text{collision}} = P_{Safe}^2 \cdot P_{Red}^2 \cdot P_{\text{Same aggregation}} \cdot P_{\text{Same Dest edge}}
\end{equation}

Substituting the terms:
\begin{equation}
    P_{\text{collision}} = \left( P_{Red} \right)^2 \cdot \left[ \left( 1 - \frac{2 \cdot T_{IPG}}{T_{SERVICE}} \right)^{\frac{\k}{2} - 1} \right]^2 \cdot \frac{2}{\k} \cdot \frac{\frac{\k}{2}-1}{\frac{\k^3}{4}-1-\frac{\k}{2}}
\end{equation}

The expected number of collisions:
\begin{equation}
    E_{\text{collision}} = \binom{\n}{2} P_{\text{collision}} = \frac{\n(\n-1)}{2} \cdot P_{\text{collision}}
\end{equation}

\mypar{Note on Independence Assumption}
The derivation above assumes statistical independence between the four categories, effectively treating the joint probability as the product of individual probabilities:
\[ P(A \cap B \cap C \cap D) \approx P(A) \cdot P(B) \cdot P(C) \cdot P(D) \]

Strictly speaking, slight dependencies exist between these events (particularly due to the \textit{sampling without replacement} nature of the Red and Destination edge logic). However, as the network size ($\n$) increases, the statistical dependence between any two specific flows weakens significantly. Consequently, this approximation becomes asymptotically accurate for large $\n$, yielding analytical results that closely match the empirical data observed in simulations.

\mypar{Simple Round Robin Synchronization}
For the Simple Round Robin routing algorithm, the calculation remains identical to the Adaptive Routing (AR) model described above, with one key exception: the second requirement, \textit{"Safe Flows"}, is not applicable.

Since RR distributes packets deterministically and is not influenced by momentary queue occupancy (unlike AR, which relies on queue states), the temporal constraint regarding $T_{IPG}$ and $T_{SERVICE}$ is irrelevant. Consequently, we are left with only three conditions:
\begin{enumerate}
    \item Red Flows
    \item Same aggregation
    \item Same Destination edge
\end{enumerate}

Mathematically, this implies that for RR, the probability of being "Safe" is effectively 1 ($P_{Safe} = 1$). The collision probability equation for RR becomes:
\begin{equation}
    P_{\text{collision, RR}} = P_{Red}^2 \cdot P_{\text{Same aggregation}} \cdot P_{\text{Same Dest edge}}
\end{equation}

Since $P_{Safe} < 1$ in the AR model, removing this term results in a higher overall probability for collision pair formation:
\begin{equation}
    P_{\text{collision, RR}} > P_{\text{collision, AR}}
\end{equation}

This theoretical conclusion aligns with the simulation results, which consistently show a higher frequency of collisions for Simple Round Robin compared to \jsq.

\begin{figure}
    \centering
    \begin{subfigure}[t]{0.48\columnwidth}
        \centering
        \includegraphics[width=\linewidth]{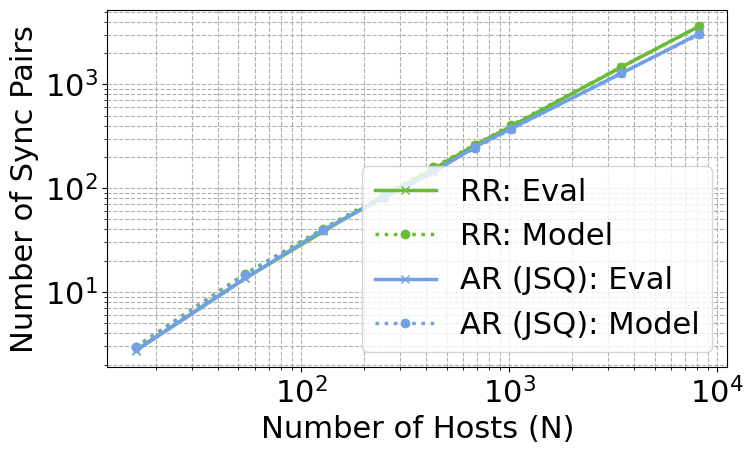}
        \caption{Synchronized pairs}
        \label{fig:pairs}
    \end{subfigure}
    \hfill
    \begin{subfigure}[t]{0.48\columnwidth}
        \centering
        \includegraphics[width=\linewidth]{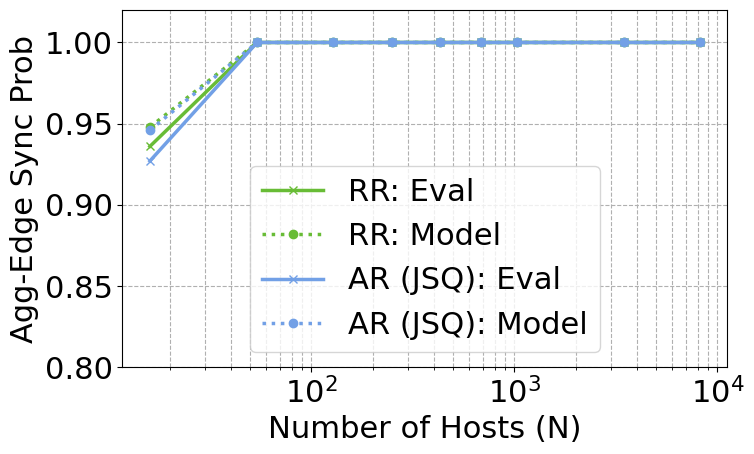}
        \caption{Aggregation-edge synchronization}
        \label{fig:prob2}
    \end{subfigure}    
    \caption{Synchronization mode for aggregation-edge links
    }
    \label{fig:ARsync}
\end{figure}

\end{proof}

\section{Proof of \cref{thm:sqrt}: Square-Root Scaling} \label{sec:sqrt}

\begin{proof}[Proof of \cref{thm:sqrt}] 
We focus again on the queue of an uplink port at an edge switch without internal traffic. The edge switch receives the sum of $\k/2$ northbound periodic flows of period $T$, and each packet enters our queue of interest with probability $1/(\k/2)$, assuming uniform hashing. Thus, the arrival rate equals the service rate, and arrivals are the sum of independent thinned periodic processes. We are at a critical loading, where the queue size behaves as a random walk with zero drift that can be modeled by a reflected Brownian motion. Specifically, within each period $T$, the number of arrived packets follows a sum of $\k/2$ Bernoulli variables, and its variance is $1-\frac{1}{\k/2}$. The queue size growth at each uplink port can be modeled as
\be
Q(m) \approx \sqrt{1-\frac{1}{\k/2}} \cdot \sqrt{\frac{2}{\pi}\cdot m},
\ee
where the first part goes to 1 as the switch size increases, and the second part shows the square-root scaling property of the random walk. 

We showed that if an edge switch does not have internal traffic, then the queue size growth at its uplink ports satisfies $Q(m) = \Omega \para{\sqrt{\m}}$ as the $m \th$ packet arrives. We now need to show that the probability that an edge switch will not have internal traffic is bounded below by a constant. 

To do so, we use \cref{eq:northbound}.
By the Weierstrass product inequality, which states that for $0 \le x_i < 1$, $\prod_{i} (1 - x_i) \geq 1 - \sum_{i} x_i$, we get 
\begin{equation}
    P_{\text{northbound}} = \prod_{i=0}^{\k/2-1} \para{1-\frac{\k/2-1}{\n-1-i}} \geq 1 - \sum_{i=0}^{k/2-1} \frac{k/2-1}{\n-1-i}
\end{equation}
The denominator is minimized at the last index, thus
\begin{equation}
    P_{\text{northbound}} \geq 1 - \sum_{i=0}^{k/2-1} \frac{k/2-1}{\n-\k/2} = 1- \frac{\frac{k}{2}(\frac{k}{2}-1)}{\n - \frac{k}{2}}
\end{equation}
Substituting $\n = k^3/4$:
\begin{equation}
    P_{\text{northbound}} \geq 1 - \frac{\frac{k^2}{4} - \frac{k}{2}}{\frac{k^3}{4} - \frac{k}{2}} = 1 - \frac{k-2}{k^2-2}.
\end{equation}
For any fat tree with $\k\geq 4,$ this last expression is larger than a constant (specifically its value of  $6/7$ for $\k=4$). Therefore the result follows.
\end{proof}

\section{Proof of \hdr Queueing Model (\cref{thm:DRB})} \label{sec:DRB}

\begin{proof}[Proof of \cref{thm:DRB}] 
We want to model the queueing behavior of \hdr when using a random permutation. First, at the edge switch uplinks, we saw that we get a sum of periodic flows with random phases. However, since the permutation is random, not all of the flows are necessarily northbound. Thus, the load $\rho$ may be arbitrary. Instead, we model it using its expected value. Any of the $\k/2$ source hosts under an edge switch sends to another flow internal to the same edge switch with probability $(\k/2-1)/(\n-1)$, therefore the average number of northbound flows is
\begin{equation}
    \frac{\k}{2}\cdot\para{1-\frac{\k/2-1}{\n-1}}
\end{equation}
Each such flow has a period of $\k/2\cdot T$, where $T$ is the packet transmission time (including inter-frame gap). We model this system as $N\cdot D/D/1$ with $\rho<1$. We model the superposition of periodic arrivals as a Gaussian process as $\k$ increases~\cite{addie2002approximation}, together with a negative drift since $\rho<1$. Since the queue size cannot be negative, we use a truncated normal distribution.

In the same way, for aggregation-to-core links, we neglect intra-pod traffic, and get the following expected number of inter-pod flows per source pod:
\begin{equation}
    \para{\frac{\k}{2}}^2\cdot\para{1-\frac{\para{\frac{\k}{2}}^2-1}{\n-1}}
\end{equation}
We then apply the same model. Note that the fit is not exact anymore, as traffic coming from the edge switches loses its periodicity and therefore it is important to emphasize that \textit{the ND/D/1 model is only an approximation}. Downlinks are an even coarser approximation, since each layer makes traffic less periodic.
\end{proof}

\begin{figure}
\centering
\begin{subfigure}[]{0.9\linewidth}
    \includegraphics[width=\linewidth]{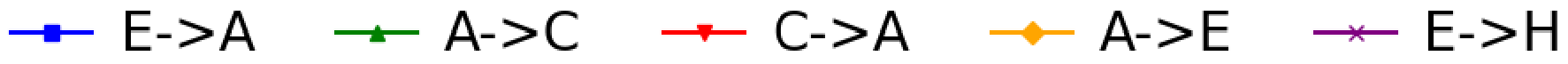}
\end{subfigure}  
\par
\begin{subfigure}{0.7\linewidth}
	\includegraphics[width=\textwidth]{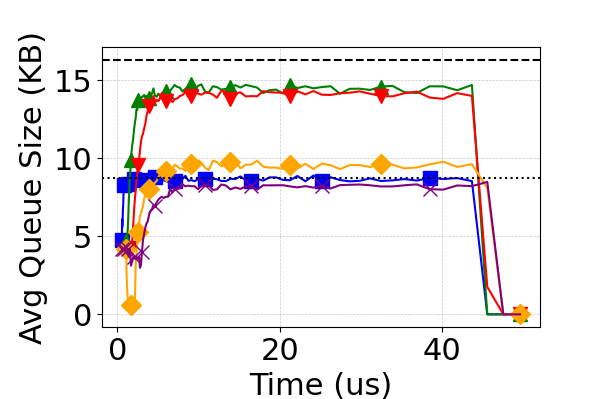}
\end{subfigure}
\caption{Queue models for \hdr. The average queue size at each of the 5 switch layers is compared against queueing models (dotted and dashed horizontal lined). }
\label{fig:drbmodel}
\end{figure}

\cref{fig:drbmodel} uses $\n=1,024$ hosts with a $4\unit{MB}$ message to show the behavior of \hdr at scale. It shows the average queue sizes at each of the 5 switch queueing layers: (i)~edge to aggregation ($E\to A$),   
(ii)~aggregation to core ($A\to C$),   
(iii)~core to aggregation ($C\to A$), 
(iv)~aggregation to edge ($A\to E$) and
(i)~edge to host ($E\to H$),  
It illustrates that the approximation is very close for the first E$\to$A layer. The model then is a bit pessimistic for the two core layers. It has three main sources of errors: (i)~the traffic periodicity is broken by queueing at successive switches, (ii)~the load varies and is different at each pod, and 
(iii)~we only use a fluid approximation. Finally, the two downlink edge layers are relatively close to the model as well despite the loss of periodicity.

\section{Proofs and Details of \name's Properties} \label{sec:ofan}

\subsection{Correctness of Consolidation}

\mypar{Load-balancing invariant} We start by formally formulating the goal of a switch \dr in a network without link failures. 
It is clear that we want a perfect LB locally among uplink ports. But we also want to make sure that any source edge switch $E$ load-balances traffic as equally as possible across the links in the destination pods. We want to formulate a target ideal invariant that this load balancing should satisfy. However, we need to be careful in the formulation. The traffic that $E$ load-balances depends on the destinations of the flows that cross $E$. These destinations depend on the arbitrary permutation pattern, and are therefore not uniform. For example, $E$ cannot send the same amount of traffic to all edge switches in the network. Instead, we condition the invariant on the destination switch.

\begin{invariant}[Load-balancing invariant] \label{inv:balance}
Given any source switch $S$ in a source pod and any destination switch $D$ in a destination pod, $S$ should send traffic at the same rate through all of the southbound links to $D$ that are reachable from $S$.
\end{invariant}

The invariant implies a perfect load balancing at downlinks:

\begin{property}\label{prop:inv}
If \inv{inv:balance} holds, then all southbound links to any switch have equal load. 
\end{property}
\begin{proof}%
    The load on each down link is the sum of the rates of all possible source edge switches, for example. Therefore, the sums will be equal for all down links to a given switch, since their components from each source edge are equal.
\end{proof}

\mypar{Consolidation} To achieve the above target invariant, we use \textit{consolidation into virtual flows}. The key idea is that each source switch consolidates into a single shared virtual flow all of its northbound packets %
destined to a shared  destination $D$. Then, it load balances these packets across all of its uplink ports. However, we still need to carefully define the level at which to consolidate flows. 

\mypar{Mandatory waypoints} Consider a source aggregation switch $A_S$. A destination host $H_D$ that is reachable through its uplinks is necessarily in a different pod. Moreover, a property of the three-level fat tree is that when sending traffic from $A_S$ to $H_D$, any packet \textit{must} reach a given destination edge $E_D$ and a given destination aggregation switch $A_D$ in the destination pod, and cannot reach other edge or aggregation switches. For example, the third aggregation switch of the source pod can only send traffic destined to $H_D$ through the third aggregation switch $A_D$ of the destination pod, then through the edge switch $E_D$ that is connected to $H_D$. We can think of $A_D$ and $E_D$ as \textit{mandatory waypoints}. They stand in contrast with the many possible core switch waypoints that could be used to reach $H_D$.

As a consequence, all of the traffic going through $A_S$ and destined to $H_D$ could be consolidated into a single virtual flow using several consolidation levels: (i)~per destination host $H_D$, (ii)~per destination edge $E_D$, and (iii)~per destination aggregation $A_D$. The following result shows that they can all provide a first performance guarantee.

\begin{theorem}\label{thm:S}
Source aggregation switch $A_S$ satisfies \inv{inv:balance} whether it consolidates traffic per destination host, edge or aggregation switch.
\end{theorem}
\begin{proof}%
    No matter the aggregation level, for any destination belonging to this aggregation level in the destination pod, $S_S$ always sends traffic at the same rate to all cores that it is connected to. Thus $S_D$, the symmetric aggregation in the destination pod that is also connected to all these cores, also receives the same rate from all these cores. Therefore, it satisfies the invariant. In addition, $S_S$ can only reach the links $S_D \to L_D$ and $L_D \to L_D$, and no other down links to $L_D$ or $H_D$. Thus, they trivially satisfy the invariant.
\end{proof}

Consider now a source edge switch $E_S$. Its mandatory waypoints for a given destination host $H_D$ are only $H_D$ and $E_D$. Since it can reach $H_D$ using several aggregation switches, none is mandatory. Thus, $E_S$ can consolidate traffic (1)~per destination host $H_D$ and (2)~per destination edge $E_D$. We find that both work out:

\begin{theorem}\label{thm:both}
No matter the consolidation level, if both the source edge switch $E_S$ and the source aggregation switch $A_S$ consolidate traffic, then both satisfy \inv{inv:balance}.
\end{theorem}
\begin{proof}%
    We structure the proof based on the destination type.
    
    (1) For any destination aggregation $S_D$, source edge $L_S$ only has a single source aggregation waypoint $S_S$ ($S_S=S_D$ if the destination is in the same pod, and it is the symmetric aggregation of $S_D$ in the source pod if they are in distinct pods). Therefore, all of its traffic will be sent to $S_S$, which satisfies the invariant for $S_D$ by \thm{thm:S}.

    (2) For any destination edge $L_D$, source edge $L_S$ equally splits all of its traffic across all of the aggregations in its pod, sending the same rate $R$ to each. In turn, the rate from any source aggregation $S_S$ is directly transferred to the symmetric destination aggregation $S_D$ (whether $S_D$ is in a different pod, or in the same pod with $S_D=S_S$). Thus destination edge $L_D$ will receive the same rate $R$ from all its uplink aggregations.

    (3) Finally, destination host $H_D$ is only connected through a single link to the network, so the invariant holds trivially.
\end{proof}

Since it is clear that pointers also load-balance perfectly locally among upling ports, the above theorem means that LB achieves its dual goals both at uplinks and at downlinks.

\mypar{Minimizing pointers} We saw that each switch can consolidate using several consolidation levels. However, \textit{the higher the consolidation level, the smoother the load balancing}, as the source switch needs to hold fewer pointers and can coordinate information across more flows. Therefore, \name always chooses the highest and largest consolidation level. 

\subsection{Scaling Impact \vs \hdr}

\mypar{Scaling impact on pointer numbers} Let's illustrate how consolidation improves \sdr (\name) \vs \hdr. Consider a random permutation with inter-pod traffic only. In \name, aggregation switches consolidate traffic per destination aggregation switch. Thus, a destination aggregation switch can only receive up to $\k-1$ different virtual flows, since there are $\k-1$ different aggregation switches. However, in \hdr, they  virtually consolidate traffic per destination host, hence there are $\k^2/4$ flows, corresponding to the $\k^2/4$ hosts in the destination pod. Thus, the consolidation gain in the number of pointers is at least $\k^2/(4(\k-1))>\k/4,$ increasing with the network size. 

\begin{figure}
    \centering
    \includegraphics[width=0.55\linewidth]{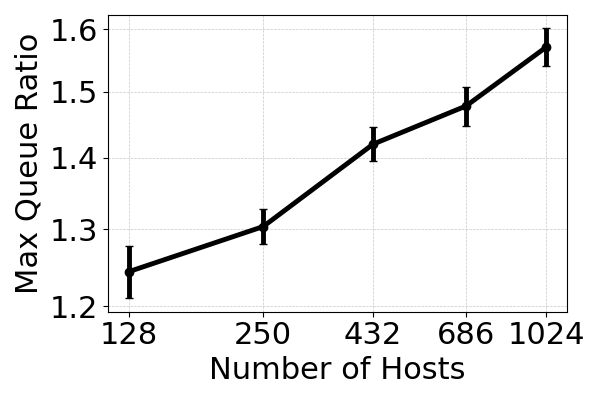}
    \caption{Ratio of the maximum queue size of \sdr (Ofan) by that of \hdr as a function of the number of hosts. 
    }
    \label{fig:ratio}
    \vspace{-1em}
\end{figure}

\mypar{Scaling impact on queue sizes} To illustrate how consolidation improves \sdr (\name) \vs \hdr, consider a random permutation with inter-pod traffic only. \Cref{fig:ratio} plots the ratio of the maximum queue size of \name by that of \hdr as a function of network size, with 30 runs per size. The ratio increases with the network size, \ie \name outperforms even more as we scale the network. 

\subsection{Model with Bounded Average Queue Sizes }

\mypar{Proof of  \cref{thm:ofan}} We are finally able to build a model for the queue size under \name.
\begin{proof}[Proof of \cref{thm:ofan}] 
The model follows similar lines for \name as for \hdr. However, since we consolidate pointers, we   evaluate the number of distinct pointers, and model as if all pointers carried the same number of periodic flows, which tends to underestimate the final queue length.

We start with the edge model. We want to determine the expected number of pointers at the edge switch. A single source edge switch comprises $\k/2$ flows (balls), and these flows can be destined to $\k^2/2$ edge switches (bins). The expected number of distinct destination edge switches is:
$$ \frac{\k^2}{2}\cdot \para{ 1 - \left( 1 - \frac{1}{\k^2/2} \right)^{\k/2} }. $$
Using Taylor expansion, we find that this is about $k/2-1/4.$

Likewise, at each aggregation switch, the number of northbound pointers is at most $\k-1$, as there are $k-1$ other pods and therefore other aggregation switches that can be destinations. The expected number of pointers also converges to $\k-1$ as $\k$ grows. Thus, we obtain a new model, which neglects many aspects of the algorithm that can be too complex to capture, and in particular the variability in the periods of each virtual flow. Again, \textit{the ND/D/1 model is only an approximation}.
\end{proof}

\begin{figure}
\centering
\begin{subfigure}[]{0.9\linewidth}
    \includegraphics[width=\linewidth]{fig/lmodel.png}
\end{subfigure}  
\par
\begin{subfigure}[]{0.7\linewidth}
	\includegraphics[width=\textwidth]{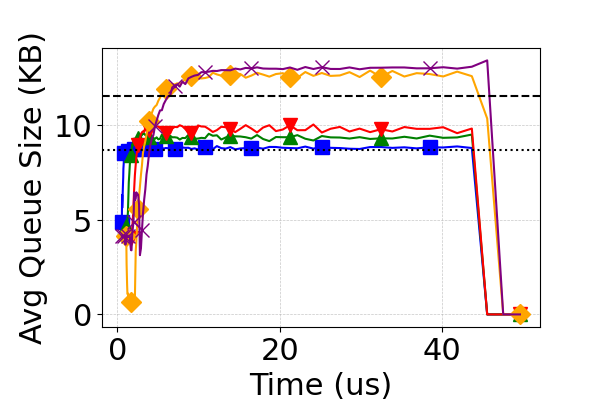}  
\end{subfigure}
\caption{Queue models for \sdr (\name). The average queue size at each of the 5 switch layers is compared against queueing models (dotted and dashed horizontal lined). }
\label{fig:ofanmodel}
\end{figure}

\Cref{fig:ofanmodel} illustrates the model for \name under the same conditions. We can see that the edge model holds well (dotted line). However, as expected, the models beyond have a slightly poorer fit. In particular, the downlink queues are  harder to model exactly.

\subsection{Advanced Implementation Considerations}

\mypar{W-ECMP} We want to make \name handle link failures, once the information has propagated at steady state. We leverage W-ECMP, which is already used in current switches. 
For each destination switch, we obtain the raw weights associated at the source switch with each port. Then, we divide by the greatest common divisor to obtain simplified weights. Next, we randomly shuffle the port order, as described above. Finally, we extract an Interleaved Weighted Round-Robin (IWRR) schedule~\cite{katevenis1991weighted,tabatabaee2021interleaved} that the pointers can go through in an efficient manner, either using direct increment or a schedule table.

\mypar{Schedule example in asymmetric network} Assume that an edge switch has 4 uplink ports $\brac{P_1,P_2,P_3,P_4}$ of W-ECMP weights $\brac{1600,1600,1600,800} \unit{Gbps}$. It first simplifies them into $\brac{2,2,2,1}$ (by dividing by the greatest common divisor). Then the port order can be reshuffled into $\brac{P_3,P_2,P_4,P_1}$, yielding for example IWRR schedule $\brac{P_3,P_2,P_4,P_1,P_3,P_2,P_1}$, where $P_4$ appears twice less often than the other ports. 

\mypar{Scalability} Assume $\k=64$ and $\n \approx 65\unit{\k}$ hosts. Each source edge switch needs to keep schedules for all of its destination edge switches, \ie up to $2,047$ schedules. This can get complex to implement and cache. 
We exploit a property intrinsic to a 3-level fat-tree: in a datacenter network with single links between switches of uniform capacity $\C$, whenever both weights are non-zero (\ie there is reachability), then (i)~the W-ECMP weight between port $i$ of source edge switch, $E_S$, and destination edge switch, $E_D$, is proportional to (ii)~the number of distinct paths between the $i\th$  aggregation switch, $A_S$, of the source pod and the $i\th$  aggregation switch, $A_D$, of the destination pod~\cite{ietf-bess-ebgp-dmz-08}. Thus, we can define a weight \textit{per destination pod}, such that the weight per destination edge switch equals this per-pod weight if it's reachable, and zero otherwise. As a result, we can only keep $k$ shared per-destination-pod schedules, a $k/2\times$ reduction ($32\times$ for our $k=64$ example), and the many pointers per (destination switch, packet type) will rotate through these shared schedules, simply double-checking FIB reachability information each time they increment and pick a new port.

\mypar{All-to-All with failure} 
We found a corner case in our simulations of \name, when testing traffic with all of the following characteristics: (i)~all-to-all, (ii)~without link failure in the network,  (iii)~sending at full link rate without congestion control, and (iv)~having a sudden link failure in the core. In such traffic, the number of source hosts per pod equals the number of cores, so each flow that exits the pod becomes anchored to a unique core in the round-robin rotation. If a link to the core suddenly fails, then all of its packets will get lost. In practice, we expect the CCA to catch this and slow down some rate within a few RTTs, losing the synchronization, as it affects a single flow and not all flows.

\mypar{Implementation variants} We also implemented two  variants to \name. First, in a network without link failures, we found that if (i)~the first pointer is not picked at random, but rather chooses the shortest queue at the arrival of the first packet, then (ii)~cycles in an unshuffled round-robin order, then performance is slightly improved. However, we lose the randomization property. 

Second, in a network with link failures, if we only use FIB reachability information without the W-ECMP weights, we found that performance slightly worsens, but it still outperforms existing algorithms. This FIB-based version can be a simpler alternative to implement.

\section{Proof of \cref{thm:pkt}: Optimal Packet Size} \label{sec:pkt}

\begin{proof}[Proof of \cref{thm:pkt}] 
The CCT can be modeled as the sum of (i)~the time to send the message, (ii)~the time to go through the queues, and (iii)~the zero-load RTT. Only the first two components depend on the packet size $P$, so let's focus on them. 

(i)~The time to send the message is
\be \frac{D}{P-H}\cdot \frac{P}{\C},
\ee
since we need $\frac{D}{P-H}$ packets to send a message of $D$ bytes using $P-H$ payload per packet, and each packet of size $P$ that is sent at rate $\C$ takes time $\frac{P}{\C}$. 

(ii)~The time to go through the queues is
\be \alpha \cdot \cdot \frac{P}{\C}.
\ee
by assumption. 

Adding the two components and dividing by constant $\C$, 
we want to find:
\be \min\quad \sbrac{P\cdot \para{\frac{D}{P-H}+\alpha}}.
\ee
Differentiating with respect to $P$, we get 
\be
\frac{D\cdot H}{\para{P-H}^2}+\alpha.
\ee
Zeroing the derivative yields the result. 
\end{proof}

Note that \cref{fig:pktsize} assumes as a first approximation $H=82\unit{B}$, accounting for the header and inter-frame gap components, and $\alpha=10$ data packets.

\section{Swift's Settings}\label{sec:swift}

In our evaluations with the MSwift CCA, we use LTCP's and MSwift's default settings. However, Swift needs tuning. 

Swift's target delay parameters allow the maximum congestion window to adapt to the sum of a fixed bandwidth-delay product (BDP) component 
and a small variable queueing component. They assume a maximum of 100 packets, given $50\unit{Gbps}$ and $100\unit{Gbps}$ links. These numbers need to be adjusted to our network with $800\unit{Gbps}$ links links and a $6.25\unit{us}$ zero-delay RTT (including propagation and serialization delays). The high line rates mean that the BDP alone already reaches 150 packets on the line. 
In the original paper~\cite{swift}, Swift uses target delays of
\begin{itemize}
    \item $25\us$ for  $50\unit{Gbps}$ links, translating to a queueing component of $(25-6.25)\cdot 50\cdot10^9 /8=117\unit{KB}$; and 
    \item $50\us$ for  $100\unit{Gbps}$ links, translating to $109\unit{KB}$.
\end{itemize}
We use the average of $113\unit{KB}$. The target is proportional to the number of hops through Swift's ${h}$ parameter. When sending at line rate, the initial congestion window is set to this target.

\end{document}